\documentclass{JHEP3}
\usepackage{amssymb}
\usepackage{amsfonts}
\usepackage{amsbsy}
\usepackage{amsmath}
\usepackage{amsthm}
\usepackage{graphicx}
\usepackage{epstopdf}
\usepackage[vcentermath]{youngtab}
\usepackage{multirow}

\def\Z{\mathbb{Z}}
\def\F{\mathbb{F}}
\def\C{\mathbb{C}}
\def\G{\mathcal{G}}
\def\L{\mathcal{L}}

\def\N{\mathcal{N}}
\def\P{\mathbb{P}}
\def\Pp{\mathbb{P}}

\def\talpha{\tilde{\alpha}}
\def\nhv{n_h-n_v}
\def\n3a{t}
\def\Tr{{\mathrm{Tr}}}
\def\tr{{\mathrm{tr}}}
\def\ord{\mathrm{ord}}
\def\O{\mathcal{O}}

\title{Mapping 6D ${\cal N} = 1$
supergravities to F-theory}

\author{Vijay Kumar$^1$, David R.  Morrison$^2$ and Washington Taylor$^1$\\
$^1$Center for Theoretical Physics\\
Department of Physics\\
Massachusetts Institute of Technology\\
Cambridge, MA 02139, USA\\
\\
$^2$Departments of Mathematics and  
Physics\\ University of California\\ Santa Barbara, CA 93106, USA\\
\\
{\tt vijayk} {\rm at} {\tt mit.edu},
{\tt drm} {\rm at} {\tt math.ucsb.edu},
{\tt wati} {\rm at} {\tt mit.edu}
}

\preprint{MIT-CTP-4087, UCSB-Math-2009-27}

\abstract{We develop a systematic framework for realizing general
anomaly-free chiral 6D supergravity theories in F-theory.  We focus on
6D $(1, 0)$ models with one tensor multiplet whose gauge group is a
product of simple factors (modulo a finite abelian group) with matter
in arbitrary representations.  Such theories can be decomposed into
{\it blocks} associated with the simple factors in the gauge group;
each block depends only on the group factor and the matter charged
under it.  All 6D chiral supergravity models can be constructed by gluing such
blocks together in accordance with constraints from anomalies.
Associating a geometric structure to each block gives a dictionary for
translating a supergravity model into a set of topological data for an
F-theory construction.  We construct the dictionary of F-theory
divisors explicitly for some simple gauge group factors and associated
matter representations.  Using these building blocks we analyze a
variety of models.  We identify some 6D supergravity models which do
not map to integral F-theory divisors, possibly indicating quantum
inconsistency of these 6D theories.  }

\begin{document}

%\input{1}
%--------------------------------
\section{Introduction}

String theory appears to provide a framework in which gravity can be
consistently coupled to many different low-energy field theories in
different dimensions.  The problem of understanding precisely which
low-energy gravity theories admit a UV completion, and which can be
realized in string theory, is a longstanding challenge.  Many different
string constructions exist, which have been shown to give a variety of
low-energy theories through compactifications of perturbative string
theory or M/F-theory.  In four space-time dimensions, while there are many
string constructions, giving a rich variety of field theory
models coupled to gravity, there is no general understanding as yet of
which gravity theories admit a UV completion and which do not.  In six
dimensions, however, we may be closer to developing a systematic
understanding of the set of allowed low-energy theories and their UV
completions through string theory.  For chiral (1, 0) supersymmetric
theories in six dimensions, cancellation of gravitational, gauge, and
mixed anomalies give extremely strong constraints on the set of
possible consistent models \cite{gsw}.  In \cite{finite}, it was shown that (with
restrictions to nonabelian gauge group structure and one tensor
multiplet) the number of possible distinct combinations of gauge
groups and matter representations appearing in such
models is finite.  In
\cite{universality}, it was conjectured that all consistent models of
this type have realizations in string theory.  The goal of this paper
is to connect the set of allowed chiral 6D supergravity theories to 
their string realizations by developing a systematic
approach to realizing these theories in F-theory.

In a general 6D supergravity theory, the gauge group can be decomposed
into a product of simple factors modulo a finite abelian group (${\cal
G} = (G_1 \times \cdots \times G_k)/\Gamma$) [In this paper we ignore
U(1) factors].  In \cite{finite} it was shown that when there is one
tensor multiplet, the anomaly cancellation conditions in 6D
independently constrain each nonabelian factor $G_i$ in the gauge
group, along with the associated matter representations, into a finite
number of distinct ``building blocks''.  Each building block makes a
contribution to the overall gravitational anomaly $\nhv = 244$, where
$n_h, n_v$ respectively are the numbers of hyper and vector multiplets
in the theory.  An arbitrary model can be constructed by combining
these building blocks to saturate the gravitational anomaly (with
neutral hypermultiplets added as needed).  The basic idea of the
approach we take in this paper is to construct a dictionary between
these building blocks of anomaly-free 6D theories and geometric
structures in F-theory.  F-theory \cite{F-theory} is a framework for
constructing type IIB string vacua where the axio-dilaton varies over
the internal space.  The nonperturbative $SL(2,\Z)$ symmetry of type
IIB is geometrized in F-theory as the modular group of a fictitious
$T^2$ fibered (holomorphically) over the internal space.  F-theory on
elliptically fibered 3-folds gives rise to a large class of 6D
theories with $(1,0)$ supersymmetry \cite{Morrison-Vafa,
Bershadsky-all}.  The low-energy theory has one tensor multiplet when
the base of the elliptic fibration is a Hirzebruch surface $\F_m$;
this is the case we will consider in this paper.  We develop a
dictionary in which each supergravity building block is associated
with a geometric structure in F-theory given by a divisor class on the
$\F_m$ base of the elliptic fibration.  Then, the construction of an
F-theory model associated with a given anomaly-free 6D model proceeds
by simply combining the divisors on the F-theory side associated with
the building blocks on the supergravity side.  The connection between
the anomaly cancellation conditions in 6D and the topological
constraints on an F-theory construction were analyzed in \cite{Sadov,
Grassi-Morrison, Grassi-Morrison-2}.  In those papers, a detailed
analysis is given of the F-theory structure associated with specific
matter representations in the associated supergravity theory.  In this
paper we combine the results of that analysis with the block
construction of supergravity theories and an explicit map from
supergravity blocks to F-theory divisors to give a complete picture of
the correspondence between 6D supergravity theories and F-theory
models.  This correspondence has potential not only to help in
understanding the string realization of various supergravity theories
in 6D (and perhaps eventually in 4D), but also to assist in
understanding the range of geometric singularities possible in
F-theory. 

In this paper we focus initially on theories with gauge group
constructed from products of simple factors $SU(N)$.  This provides a
clean and fairly simple illustration of the general ideas just
described.  A similar analysis is also possible for the other
classical groups $SO(N)$ and $Sp(N)$, and the exceptional groups $E_6,
E_7, E_8, F_4, G_2$.  We give some simple examples of these other
groups, leaving a systematic analysis of F-theory geometry associated
with arbitrary gauge group and matter representations for future work.
We identify some situations in which the map to F-theory violates an
integrality condition on divisors in the base of the F-theory
construction, so that apparently no F-theory model exists
corresponding to these supergravity theories.  We speculate on
possible associated integrality constraints on the low-energy
theories.

In Section \ref{sec:supergravity} we review the structure of anomalies
in 6D (1, 0) supergravity theories.  We summarize the results of
\cite{finite} showing that the number of consistent theories with one
tensor multiplet is finite, and elaborate on the construction of
models from building blocks associated with factors in the gauge
group.  We explicitly describe the allowed factors with gauge group
$SU(N)$ and matter in the fundamental and antisymmetric tensor
representations, which form a simple example of the general framework
presented here.  In Section \ref{sec:F-theory} we review the relevant
basic structures in F-theory.  We give an explicit dictionary from
$SU(N)$ supergravity building blocks to divisors in F-theory, and find
that all product group models built from these blocks in supergravity
give rise to topologically allowed combinations of divisors in
F-theory.  In Section \ref{sec:more-representations} we expand the
dictionary to include other representations of $SU(N)$ as well as some
other simple groups and representations, and describe the
corresponding structure in F-theory.  In Section \ref{sec:Weierstrass}
we discuss the problem of constructing explicit Weierstrass models
associated with the topological data given by the dictionary for a
given supergravity model.  In Section \ref{sec:summary} we summarize
some of the exceptions we have identified to the integrality of the
F-theory mapping.  We conclude in Section \ref{sec:conclusions} with a
general discussion and comments on extensions of the results described
in this paper.  Related work analyzing the interplay of
constraints imposed by anomaly cancellation and geometric constraints
in F-theory has recently appeared in \cite{Choi, Laamara}.

\section{Anomaly-free (1, 0) supergravity models in 6D}
 \label{sec:supergravity}

In this section we review the basic anomaly conditions of $(1, 0)$
supersymmetric theories coupled to gravity in six dimensions
(subsection \ref{sec:review}), and the
result of \cite{finite} showing that only a finite number of gauge
groups and matter content are possible in such theories
(subsection \ref{sec:finite}).  We then give an example of how a class of such
models can be explicitly enumerated by giving a complete
classification of all models whose gauge group is a product of $SU(N)$
factors, with matter in fundamental, antisymmetric, and bifundamental
representations (subsection \ref{sec:classification}).

\subsection{Review of anomaly conditions}
\label{sec:review}

In this subsection we give a brief review of the anomaly conditions on
6D (1, 0) supergravity theories \cite{gsw}.  A more complete review of
these conditions appears in \cite{finite}, and we mostly follow the
notation and conventions of that paper.  We repeat some of the central
equations here for convenience.

Throughout the paper
we
denote traces in the fundamental and adjoint representations by
tr, Tr respectively, using
tr$_{R}$  for all other representations $R$.
Traces of second and fourth powers of $F$ in any representation can be
expanded as
\begin{eqnarray}
\tr_R F^2 & = &  A_R \tr F^2 \label{eq:decomposition-1}\\
\tr_R F^4 & = &  B_R \tr F^4 + C_R (\tr F^2)^2 \label{eq:decomposition-2}
\end{eqnarray}
We denote the dimension of a general representation $R$ by $D_R$.

We consider theories with gauge group of the form ${\cal G} = (G_1
\times \cdots \times G_k)/\Gamma$ with $G_i$ simple (assuming no U(1)
factors) and $\Gamma$ a finite abelian group.  
The number of hypermultiplets in representation $R$ of group $i$ is
denoted $x^i_R$, and the number of bifundamental hypermultiplets
 transforming in $(R,
S)$ under $G_i, G_j$ is denoted $x^{ij}_{RS}$.

We let $n_t$, $n_h$ and $n_v$ denote the number of tensor multiplets,
hypermultiplets, and vector multiplets in our theory.
For $n_t = 1$, the anomaly has a term proportional to
\begin{equation}
I_1 = (n_h-n_v-244) {\rm Tr}\; R^4 \,,
% % \label{eq:}
\end{equation}
whose vanishing implies
\begin{equation}
n_h-n_v= 
%\sum_{i}\left[\left( \sum_{R}   n^i_R D_R^i \right) -D_{\rm
%  adjoint}^i \right]=
244
\label{eq:bound}
\end{equation} 
When 
(\ref{eq:bound}) is satisfied,
the anomaly polynomial becomes
(after rescaling so that the coefficient of
$(\tr R^2)^2$ is one)
\begin{align}
I &= (\tr R^2)^2 +\frac{1}{6} \tr R^2\sum_i \left[ \Tr F_i^2 - \sum_R x^i_{R} \tr_{R} F_i^2\right] - \frac{2}{3}\left[ \Tr F_i^4 - \sum_R x^i_{R} \tr_{R} F_i^4\right] \notag\\
& \quad +4 \sum_{i,j,R,S} x^{ij}_{RS} (\tr_{R} F_i^2)(\tr_{S}
F_j^2)
\label{eq:scaled-anomaly}
\end{align}
Anomalies can
be cancelled through the Green-Schwarz mechanism \cite{gsw, Erler}
when this polynomial can be factorized as
\begin{equation}
I = (\tr R^2 - \sum_i \alpha_i \tr F_i^2)(\tr R^2 - \sum_i \talpha_i \tr F_i^2) \label{eq:factorized-anomaly}
\end{equation}

A necessary condition for the anomaly to factorize in this fashion is
the absence of any irreducible $\tr F_i^4$ terms.  This
gives the condition
\begin{align}
\tr F_i^4 :   & \quad B^{i}_{Adj}  =  \sum_R x^i_{R} B^i_{R} \label{eq:f4-condition}
\end{align}
For groups $G_i$ which do not have an irreducible $\tr F_i^4$ term,
$B^i_R=0$ for all representations $R$ and therefore
(\ref{eq:f4-condition}) is always satisfied.  The sum in
(\ref{eq:f4-condition}) is over all hypermultiplets that transform
under any representation $R$ of $G_i$.  For example, a single
hypermultiplet that transforms in the representation $(R,S,T)$ of
$G_i\times G_j \times G_k$ contributes $\dim(S)\times \dim(T)$ to
$x^i_R$.  Note that the anomaly conditions are not sensitive to
whether a group transforms in a given representation $R$ or the
conjugate representation $\bar{R}$.  For example, in a model with
gauge group $SU(N) \times SU(M)$ with $x$ hypermultiplets in
$(N,\bar{M}) + (\bar{N}, M)$ and $y$ hypers in $(N,M)+ (\bar{N},
\bar{M})$, anomaly cancellation can only constrain the sum $x+y$.
F-theory in its usual formulation generally gives rise only to
hypermultiplets in the first category, with $y = 0$.  An F-theory
realization of models with $y \neq 0$ has not been fully developed,
though such supergravity
models certainly are possible in six dimensions, as found
for example in \cite{Kumar-Taylor}.

For a factorization of the anomaly polynomial
(\ref{eq:scaled-anomaly}) to exist, in addition to (\ref{eq:bound})
and (\ref{eq:f4-condition}), the following equations must have a
solution for real $\alpha_i, \talpha_i$
\begin{align}
\alpha_i+\talpha_i & = \frac{1}{6} \left( \sum_R x^i_{R} A_{R}^i -
A_{Adj}^i\right)  \label{eq:a-condition}\\
\alpha_i\talpha_i & = \frac{2}{3}\left( \sum_R x^i_{R} C_{R}^i -
C_{Adj}^i \right)  \label{eq:c-condition}\\
\alpha_i\talpha_j +\alpha_j\talpha_i & = 4 \sum_{R,S} x^{ij}_{RS}
A_{R}^i A_{S}^j \label{eq:bifundamental-condition}
\end{align}

\subsection{Finite number of models}
\label{sec:finite}

In \cite{finite}, it was proven that there are a finite number of
distinct gauge groups and matter representations which satisfy the
conditions (\ref{eq:bound}), (\ref{eq:f4-condition}),
(\ref{eq:a-condition}), (\ref{eq:c-condition}), and
(\ref{eq:bifundamental-condition}) when the additional condition is
imposed that all gauge kinetic terms must be positive for some value
of the dilaton.

The condition (\ref{eq:bound}) plays a key role in this proof of
finiteness.  The anomaly cancellation conditions constrain the matter
transforming under each gauge group so that the quantity $n_h-n_v$ in
general receives a positive contribution from each gauge group and
associated matter, and the construction of models compatible with
(\ref{eq:bound}) thus has the flavor of a partition problem.  (There
are cases where a single gauge group factor and associated matter
contribute a negative $\nhv$, but generally only one such factor can
appear in any model).  Because equations (\ref{eq:a-condition}),
(\ref{eq:f4-condition}) and (\ref{eq:c-condition}) all depend only
upon the numbers of fields transforming in different representations
under the gauge group factor $G_i$, we can consider solutions of these
equations as ``building blocks'', from which complete theories can be
constructed by combining building blocks, with the overall constraint
(\ref{eq:bound}) bounding the size and complexity of the possible
models which can be constructed.  In combining blocks in this fashion,
it is necessary to keep in mind that some matter transforming under a
given gauge group may also have nontrivial transformation properties
under another group, so that $\nhv$ is subadditive.  The number of
such fields transforming under multiple groups, however, is bounded by
(\ref{eq:bifundamental-condition}), so that the enumeration is still
finite.

The parameters $\alpha_i, \tilde{\alpha}_i$ which are fixed through
(\ref{eq:a-condition}), (\ref{eq:c-condition}) for each block (up to
exchanging the two values) play a key role in the structure of
consistent models.  These parameters enter the Lagrangian in the
kinetic term for the gauge field $G_i$ through
\begin{align}
\L = 
-\sum_i (\alpha_i e^\phi+\talpha_ie^{-\phi}) \tr(F_i^2)+\ldots
\label{eq:lagrangian} 
\end{align}
as shown in \cite{Sagnotti}.  Thus, if both $\alpha, \tilde{\alpha}$
are negative for some gauge group, the gauge kinetic term always has
the wrong sign; we do not consider theories with this apparent
instability.

As we will see in Section \ref{sec:F-theory}, the parameters $\alpha,
\tilde{\alpha}$ are the key to the mapping from gauge group building
blocks to F-theory.  These parameters encode the homology class of the
divisor in the base of the F-theory compactification associated with
the given gauge group component.  From the anomaly cancellation
equations, it is clear that there are various constraints on the
$\alpha_i, \tilde{\alpha}_i$ parameters for the various gauge group
components.  For example, we cannot have two gauge group factors which
both have $\alpha_i < 0$ and $\tilde{\alpha}_i > 0$, or the number of
bifundamental fields charged under these two groups would be negative
through (\ref{eq:bifundamental-condition}).  Similarly, two factors
with matter representations giving specific values of the $\alpha,
\tilde{\alpha}$'s cannot appear in the same theory unless the product
in (\ref{eq:bifundamental-condition}) computed using those values of
$\alpha, \tilde{\alpha}$ is divisible by 4.  These algebraic
constraints on supergravity blocks correspond to geometric constraints
in F-theory which we will describe in \ref{sec:F-theory}.

\subsection{Classification of $SU(N)$ models}
\label{sec:classification}

From the proof of finiteness and the block decomposition structure of
a general chiral 6D supergravity theory, in principle it should be
possible to systematically classify and enumerate all possible models,
at least when restricting to a semisimple gauge group and one tensor
multiplet.  Each model has a gauge group $\G = (G_1\times \ldots \times
G_K)/\Gamma$, and matter multiplets in any representation of $\G$.  In
classifying all 6D models, a key point is that the values of $\alpha_i,
\talpha_i$ for each factor $G_i$ depend on the matter charged under
$G_i$ alone, and not on the other factors in the gauge group.  Thus, a
complete classification of models can proceed  heuristically as follows

\begin{enumerate}

\item Classify all blocks.

For each simple group $G_i$, classify all representations $R$ and
matter multiplicities $x^i_R$, such that $\sum_R x^i_R B^i_R =
B^i_{Adj}$ and solutions to (\ref{eq:a-condition}) and
(\ref{eq:c-condition}) exist.  This gives a set of building blocks,
which can be used to build the full gauge group $\G$.  We define a
{\it block} as consisting of a gauge group $G_i$ and all the
associated charged matter representations.  The values for $\alpha_i,
\talpha_i$ are determined for a given block (up to exchange).

\item  Combine blocks.

We wish to combine blocks in all possible combinations compatible with
the anomaly conditions.  The blocks from Step 1 cannot be combined
arbitrarily; in a model with gauge group $\prod_i G_i$, for every pair
of indices $i,j$, the associated blocks can only be combined if there
is enough matter which is simultaneously charged under $G_i\times G_j$
to satisfy equation (\ref{eq:bifundamental-condition}).  This gives a
constraint on which blocks can be combined, which becomes quite strong
as the number of blocks is increased.  Thus, to construct all models
we need to classify all possible combinations of the blocks determined
in Step 1, subject to the conditions that both
(\ref{eq:bifundamental-condition}) and the gravitational anomaly
condition $\nhv=244$ are satisfied.
\end{enumerate}
Note that once the blocks have been combined, there are only finitely
many choices for the finite abelian group $\Gamma$, which is constrained
by the matter representation.  (For example, an $SU(2)$ block with
only adjoint matter could have gauge group $SU(2)/\mathbb Z_2 \cong SO(3)$,
but if there is fundamental matter it is not possible to take
$\Gamma=\mathbb Z_2$.)

This general strategy for classifying models is complicated by the
fact that even though there are only finitely many models in total,
placing a bound on the set of blocks needed in step 1 above is
nontrivial.  It is the gravitational anomaly condition $\nhv=244$,
which depends on {\it all} the matter, that ultimately enforces
finiteness.  Thus, at the level of enumerating the blocks, a block in
a given model could contribute more than 244 to $\nhv$, if another
block has a negative contribution.  Moreover, in the presence of
matter charged simultaneously under multiple groups, the contribution
to $\nhv$ from a given block is overcounted since many groups
``share'' the same hypermultiplets.

While the proof of finiteness in \cite{finite} demonstrates that a
complete enumeration of all models is in principle possible, we do not
present here a complete algorithm for efficient enumeration of all
models.  Instead, we consider a simplified class of models for which
we carry out a complete classification of models, as an example of how
the bounds from the gravitational anomaly and multiply-charged matter
fields can be used to constrain the set of possible models.  The
approach used in this simplified class could be generalized to
include most other gauge groups and matter representations, but we
leave a completely general analysis to further work.

Thus, in this paper, we implement an explicit algorithm based on the
above strategy to enumerate all models consisting of blocks with an
$SU(N)$ gauge group and associated matter in the fundamental and
antisymmetric representations.  In this section we restrict to $SU(N)$
blocks with $N > 3$, and discuss the special cases $SU(2), SU(3)$ in
Section \ref{sec:more-representations}\footnote{For
$SU(2)$, $SU(3)$, there is no quartic Casimir, and the range of models
is slightly larger both on the supergravity side and the F-theory
side.}.  We give a simple classification of all blocks of this type
(step 1), and find a lower bound for the contribution of each block to
$\nhv$ which enables us to systematically classify all models built
from these blocks (step 2), under the assumption that the only type of
matter charged under more than one gauge group factor is bifundamental
matter charged under two simple factors.  This gives a fairly simple
set of possibilities which provide a clear framework for demonstrating
the dictionary for associating blocks and complete models with
F-theory constructions, which we describe in Section
\ref{sec:F-theory}.  In Section \ref{sec:more-representations} we
describe other representations for matter charged under $SU(N)$ and
other gauge groups, and give some examples of blocks including these
structures.

\vspace{0.15in}
\noindent
{\bf Step 1:} $SU(N)$ blocks with fundamental (${\tiny\yng(1)}$) and
  antisymmetric (${\tiny\yng(1,1)}$) matter
\vspace*{0.1in}

For the fundamental and antisymmetric representations of $SU(N)$ we have

\begin{center}
\begin{tabular}{|r|c | c | c | c |}
\hline
rep.\ R & $A_R$ & $B_R$ & $C_R$ & $D_R$\\
\hline
fundamental (f) & 1 & 1 & 0 & $N$\\
antisymmetric (a) & $N -2$ &  $N -8$ & 3 & $N(N -1)/2$\\
\hline
\end{tabular}
\end{center}

For a gauge group factor $G_i = SU(N), N > 4$ with matter in only
these representations,
the $F^4$ anomaly condition (\ref{eq:f4-condition}) can be used to determine  a
relationship between the
number $f$ of fundamental representations and the number $a$ of
antisymmetric representations
\begin{equation}
f =2 N -a (N -8) \,.
\label{eq:ffa}
\end{equation}
Using this relation, a simple computation shows that
the anomaly polynomial automatically factorizes as
\begin{equation}
I = (\tr R^2 - 2 \tr F^2)(\tr R^2 -(a-2)
\tr F^2) 
% % \label{eq:}
\end{equation}
The values
of $\alpha, \tilde{\alpha}$ are therefore given by one of the two possibilities
\begin{eqnarray}
\alpha, \tilde{\alpha} & = &  2, a-2 \nonumber\\
\alpha, \tilde{\alpha} & = &  a-2, 2 \,.
\end{eqnarray}
Thus, for this sub-family of $SU(N)$ blocks with fundamental and
antisymmetric matter, we have implemented Step 1 of the algorithm
above.  Each block is specified by integers $a, N$ where $N \geq 4$,
and since $f, a \geq 0$, we have the further constraints  $2N/(N-8) \geq
a \geq 0$.

For example, consider a model where the gauge group has a single
simple factor $SU(N)$.  With various numbers $a$ of antisymmetric
representations, we find solutions which undersaturate the
gravitational anomaly condition $\nhv = 244$ up to $a = 10$, with $f =
(2-a)N + 8a$, and with $N$ ranging up to $N \leq (15, 15, 16, 18, 16,
13, 9, 6, 5, 4, 4)$ (for $a$ from 0 to 10).  The ``building blocks''
associated with these gauge groups and matter representations are
tabulated in Table~\ref{t:simple-blocks}.

Note that by
plugging 1/2 of (\ref{eq:ffa}) in to the formula (\ref{eq:bound})
for the number of matter fields, we have
\begin{equation}
\nhv = f N + a N (N -1)/2-N^2 + 1
=  N \left( f/2 + 7a/2 \right) + 1 \,.
\label{eq:fa-count}
\end{equation}
This shows that the contribution from each block to $\nhv$ is
positive, and is greater than $N f/2$.  As we now discuss, the form
(\ref{eq:fa-count}) of the contribution to $\nhv$ gives a finite bound
on the set of blocks which may enter into complete models and makes it
possible to analyze all models composed of these blocks in an
efficient fashion.
\vspace*{0.15in}

\noindent
{\bf Step 2:} Combining $SU(N)$ blocks with fundamental +
antisymmetric matter into complete models
\vspace*{0.1in}

We now wish to combine the blocks described above into all possible
models with gauge group of the form $\G = (G_1\times \ldots \times
G_K)$ (we do not explicitly carry out the analysis of possible
quotients by a discrete group $\Gamma$ here, but this could be done
systematically in a straightforward fashion).
We assume in this analysis that the only kind of multiply-charged
matter available is bifundamental matter charged under two groups
$G_i$, $G_j$.  If the number of such bifundamental fields is $x_{ij}$
then the total contribution to $\nhv$ from blocks $i, j$ is decreased
by $x_{ij} N_i N_j$.  We can subtract half this contribution from the
contribution (\ref{eq:fa-count}) from each block to $\nhv$.  This
removes at most $Nf/2$ from each block.  Thus, even with the
overcounting from bifundamentals, each block has a contribution of at
least
\begin{equation}
(\nhv)_i \geq 7 N a/2 + 1
\label{eq:an-bound}
\end{equation}
to the total gravitational anomaly.  This provides an immediate upper
bound on the set of individual blocks which can be used.  Since we
must have $(\nhv)_i \leq 244$, we have $a \leq 17$.  For $a > 0$ we
have $N \leq 486/(7a)$.  

The blocks with $a = 0$ form a special case.  For these blocks,
$(\alpha, \tilde{\alpha}) = (2, -2)$ or $(-2, 2)$.  We cannot have
more than one such block.  If we chose two blocks with the same
$\alpha, \tilde{\alpha}$, there would be a negative number of
bifundamentals from (\ref{eq:bifundamental-condition}).  And if we
choose one of each sign, then the gauge kinetic terms
(\ref{eq:lagrangian}) have opposite signs so one will be negative and
unphysical.  Thus, we can only have one block with $a = 0$.  Without
loss of generality we assume it has $(\alpha, \tilde{\alpha}) = (2,
-2)$.  We cannot have a block of this type and a block with $a = 1$,
since we would not have an acceptable number of bifundamentals between
these blocks.  Thus, if we have a block with $a = 0$, all other blocks
must have at least $a = 2$.  A similar argument shows that only one
block can have $a = 1$.  We find therefore that an efficient approach
to classifying all models is to begin by classifying all combinations
of blocks with $a > 1$, and then for each such combination to check
which blocks with $a = 0$ or $a = 1$ can be included.  While
(\ref{eq:an-bound}) does not provide a bound on $N$ when $a = 0$, if
we add the $a = 0$ block last, then $\nhv$ and/or
(\ref{eq:bifundamental-condition}) provide a strong constraint on the
$N$ allowed for the $a = 0$ block.

It is now straightforward to systematically enumerate all models built
from $SU(N)$ blocks with $a$ antisymmetric matter fields and $f$
fundamental matter fields, using (\ref{eq:an-bound}) and
(\ref{eq:bifundamental-condition}).  We can do this recursively,
starting with 1 block and continuing to $K$ blocks, adding blocks with
nonincreasing values of $a > 1$ so that
\begin{equation}
\sum_{i} (\frac{7 N_ia_i}{2}  + 1) \leq 244 ,
\label{eq:partial-condition}
\end{equation}
where at each step we only add blocks where there are enough
fundamentals in each component of the complete model to satisfy
(\ref{eq:bifundamental-condition}).  Given $K$ blocks satisfying
(\ref{eq:partial-condition}), we can then keep that combination, or
add a single block with $a = 1$ or $a = 0$.  Given all the blocks, we
then check that the total gravitational anomaly is undersaturated
\begin{equation}
\nhv \leq 244
% \label{eq:}
\end{equation}
and saturate the anomaly with neutral hypermultiplets as needed.

We have carried out this algorithm and enumerated all possible models
of this type which are consistent with all anomaly cancellation
conditions.  The number of models with $K$ blocks is tabulated in
Table~\ref{t:block-models}.  The total number of models with any
number of blocks and this gauge group and matter structure is 16,418.
In this enumeration we have restricted to $SU(N)$ blocks with $N >
3$.  In Section \ref{sec:23} we include $SU(3)$
blocks in the enumeration.
\begin{table}
\begin{center}
\begin{tabular}{||c || c|c | c|c | c | c | c | c | c ||}
\hline
$K$ & 1 & 2 & 3 & 4 & 5 & 6 & 7 & 8 & 9 \\
\hline
\# models &
88 &
1301 &
4798 &
5975 &
3202 &
882 &
152 &
19 &
1 \\
\hline
\end{tabular}
\end{center}
\caption[x]{\footnotesize Number of models with $K$ blocks, gauge
group product of $SU(N)$ factors ($N > 3$ for each factor) with matter
in fundamental and antisymmetric representations.}
\label{t:block-models}
\end{table}

As an example of a consistent theory with a product group structure
satisfying anomaly cancellation, consider the following two ``building
blocks'' associated with group factors, matter
representations, and compatible choices of $\alpha, \tilde{\alpha}$
\begin{eqnarray}
SU(4): & \hspace*{0.1in} & 2 \;{\tiny\yng(1,1)}+ 16 \;{\tiny\yng(1)},
\hspace*{0.1in} \alpha_1 = 0, \tilde{\alpha}_1 = 2  \nonumber\\
SU(5): & \hspace*{0.1in} & 4 \;{\tiny\yng(1,1)}+ 22 \;{\tiny\yng(1)} 
\hspace*{0.1in} \alpha_2 = 2, \tilde{\alpha}_2 = 2\,.
\label{eq:product-example}
\end{eqnarray}
Since we have $\alpha_1\talpha_2 +\alpha_2\talpha_1 = 4$ there is one
bifundamental hypermultiplet transforming under the $(4, 5)$
representation of the gauge group.  This uses up 5 of the fundamentals
in the $SU(4)$ and 4 of the fundamentals in the $SU(5)$.  The total
contribution to $\nhv$ from this product group is $\nhv = 167$.
This is one of the 1301 two-block models appearing in Table~\ref{t:block-models}.

As another example, consider the single model of this type with the
most blocks, appearing in the $K = 9$ column in
Table~\ref{t:block-models}.  This model has gauge group
\begin{equation}
SU(16) \times SU(4)^8
\label{eq:16-48}
\end{equation}
where the $SU(16)$ has $a = 0$ antisymmetric matter fields and each
$SU(4)$ has $a = 2$.  It follows that the $SU(16)$ block has $\alpha =
2, \tilde{\alpha} = -2$ and the other blocks have $\alpha = 0,
\tilde{\alpha} = 2$.  There are thus bifundamental fields in the $(16,
4)$ connecting the first component to each other component.  This
model has a total gravitational anomaly contribution of $\nhv = 233$,
so there are 11 neutral hypermultiplets.

We now show how all 16418 of the models classified here and
enumerated in Table~\ref{t:block-models} can be embedded in F-theory
(at least topologically).

\section{F-theory realizations of $SU(N)$ product models}
\label{sec:F-theory}

We now describe the mapping from the gauge group block construction of
consistent 6D supergravity theories to F-theory, focusing on the
simple class of models described in the previous section.  We begin
with a brief review of some basic aspects of F-theory and then
describe the map.

\subsection{Review of 6D F-theory constructions}

Compactifications of F-theory on elliptic Calabi-Yau 3-folds generate
a large class of six-dimensional theories.  Since we have restricted
our attention to models with one tensor multiplet, the base of the
elliptic fibration must be a Hirzebruch surface $\F_m$.  We briefly
review the structure of these compactifications here, for more details
see \cite{F-theory, Bershadsky-all}.

F-theory provides a geometric understanding of compactifications of
type IIB string theory where the axio-dilaton varies over the internal
space.  F-theory on an elliptically fibered Calabi-Yau $M$ with base
$B$ is a type IIB compactification on $B$, where the axio-dilaton is
identified with the complex structure of the elliptic fiber.  In our
case, $M$ is a 3-fold with base $\F_m$.  There is a codimension-one
locus in the base where the fiber degenerates; these correspond to
7-branes wrapping a complex curve.  Possible degeneration structures
are given by the Kodaira classification \cite{Kodaira}; for example, a
type I singularity along an irreducible curve $\xi$ in the base $B$ is
associated with the Dynkin diagram $A_{N-1}$, and corresponds to $N$
7-branes wrapping $\xi$.  Such a configuration generally results in an $SU(N)$
gauge group in the low-energy theory.  Similarly, all other A-D-E
gauge groups can be obtained by engineering the appropriate
degeneration on the curve $\xi$.  The set of 7-branes allowed in the
compactification is constrained by the condition that the full
manifold defined by the elliptic fibration must be Calabi-Yau, and
thus have vanishing canonical class.  Kodaira's formula expresses this
fact as a relationship between the locus of singular fibers and the
canonical class $K$ of the base.
\begin{equation}
\sum_\beta a_\beta X_\beta = -12 K \,.
 \label{eq:Kodaira-formula}
\end{equation}
Here the $X_\beta$ denote the classes of irreducible curves along
which the elliptic fibration degenerates.  The multiplicities
$a_\beta$ are determined by the singularity type of the elliptic fiber
\cite{F-theory}; for an $A_{N-1}$ singularity, the multiplicity is
$N$.  Some of the divisors $X_\beta$ correspond to curves where the
singularity in the elliptic fiber results in nonabelian gauge
symmetry; we denote the classes of such curves by $\xi_i$.  The
remaining curves do not enhance the gauge group in the low-energy
theory (singularity type $I_1$ or $II$, with multiplicity $a_\beta=1$
or $a_\beta=2$, respectively); we denote the sum of the classes of
such curves by $Y$.  Given such a decomposition of the singularity
locus, the matter content in the theory is found by studying the
detailed structure of the singularities and intersections of the
divisors $\xi_i, Y$.  The analysis of singularity types associated
with matter in various representations of the gauge group is given in
\cite{Bershadsky-all, Sadov, Katz-Vafa, Grassi-Morrison,
Grassi-Morrison-2}.  The simplest example of a matter field is when
two components of the divisor intersect.  For example, when an
$A_{N-1}$ locus, corresponding to $SU(N)$, intersects an $A_0$ locus,
the singularity type is enhanced to $A_N$ at the intersection.  This
results in matter hypermultiplets localized at the intersection, which
transform in the fundamental of $SU(N)$.  The $SU(N)$ transformation
properties of these  matter hypermultiplets are precisely the same
as if they had been obtained from the Higgsing of the adjoint of
$SU(N+1)$ \cite{Katz-Vafa}, and the Higgsing procedure is a good
informal guide to the behavior at the intersection point (although,
crucially, the gauge group itself is not actually enhanced 
to $SU(N+1)$).  Note that
while the way in which the gauge symmetry group is encoded in the
singularity structure is completely determined by the Kodaira
classification, there is as yet no complete classification of
singularity structures associated with matter representations.
Indeed, we encounter a number of exotic representations in the
classification of 6D supergravity models which should correspond to
currently unknown singularity structures on the F-theory side.  Some
examples of this type are given in section
\ref{sec:more-representations}.

We are interested in elliptic fibrations over the Hirzebruch surface
$\F_m$.  These are a family of surfaces which are $\Pp^1$ bundles over
$\Pp^1$ indexed by an integer $m\geq 0$.  A basis for the set of
divisors  is given by $D_v, D_s$, with
intersection pairings
\begin{equation}
D_v \cdot D_v = -m, \;\;\;\;\;
D_v \cdot D_s = 1, \;\;\;\;\;
D_s \cdot D_s =  0 \,.
% % \label{eq:}
\end{equation}
In terms of the fibration, $D_v$ is a section, while $D_s$ corresponds to the class of the fiber.  It is sometimes useful to work with $D_u = D_v + m D_s$, which satisfies $D_u \cdot D_u = m$.  $K$, the canonical class of $\F_m$ is given by
\begin{equation}
-K = 2D_v + (2 + m) D_s \,.
\label{eq:f-k}
\end{equation}
For $\F_m$, the effective divisors that correspond to irreducible curves are given by
\begin{equation}
D_v,\hspace*{0.1in} 
a D_u + bD_s, \; a,b \geq 0 \,.
\label{eq:f-effective}
\end{equation}

We are interested in constructing F-theory compactifications on Calabi-Yau 3-folds elliptically fibered over $\F_m$, for models with gauge groups $\prod_i SU(N_i)$, $N_i \geq 4$.  For this purpose, the singular locus must contain divisors $\xi_i$ corresponding to irreducible curves (\ref{eq:f-effective}), satisfying the Kodaira formula
\begin{equation}
24 D_v + (12m + 24) D_s = \sum_{i}N_i \xi_i + Y
\label{eq:Kodaira-formula-2}
\end{equation}
We now proceed to identify such models by mapping solutions of the
anomaly cancellation conditions into F-theory.

\begin{table}
\begin{center}
\begin{tabular}{||c | c|c | c|c || c | c | c ||}
\hline
\hline
$a$ & $f$ & max $N$ & $\alpha$ & $\tilde{\alpha}$ &
$\F_0$ & $\F_1$ & $\F_2$\\
\hline
0 & $2 N$ & 15 & 2 & -2 & & & $D_v$\\
\hline
1 & $N + 8$ & 15 & 2 & -1 & & $D_v$ &\\
\hline
2 & $16$ &  16 & 2 &  0 & $D_v$ & &\\
 &  &   & 0 &  2 & $D_s$ &$D_s$ &$D_s$\\
\hline
3 & $ - N + 24$ &  18 & 2 & 1 & & $D_v + D_s$ &\\
\hline
4 & $ - 2N + 32$ &  16 & 2 & 2 & $D_v + D_s$& & $D_v + 2D_s$\\
\hline
5 & $ - 3N + 40$ &  13 & 2 & 3 & & $D_v + 2D_s$ &\\
\hline
6 & $ - 4N + 48$ &  9 & 2 & 4 & $D_v + 2D_s$& & $D_v + 3D_s$\\
 & &   & 4 & 2 & $2D_v + D_s$&  $2 (D_v + D_s)$& \\
\hline
7 & $ - 5N + 56$ &  6 & 2 & 5 & & $D_v + 3D_s$ &\\
\hline
8 & $ - 6N + 64$ &  5 & 2 & 6 & $D_v + 3D_s$& & $D_v + 4D_s$\\
 &  &  & 6 & 2 & $3D_v + D_s$& & \\
\hline
9 & $ - 7N + 72$ &  4 & 2 & 7 & & $D_v + 4D_s$ &\\
\hline
10 & $ - 8N + 80$ &   4 & 2 & 8 & $D_v + 4D_s$& & $D_v + 5D_s$\\
 & &   &  8 & 2 & $4D_v + D_s$&  & \\
\hline
\end{tabular}
\end{center}
\caption[x]{\footnotesize  Building blocks associated with gauge group
factors $SU(N)$ having $a$ 2-index antisymmetric representations, up
to $a = 10$.  For
each block, number of fundamental representations given as function of
$N$.  Maximum value of $N$ is indicated such that $\nhv \leq 244 $,
corresponding to
constraint on single block.  (Larger $N$ for given $a$ can appear in multi-block models.)
Possible values of $\alpha, \tilde{\alpha}$ and associated divisors in
$\F_m$ are given for $m = 0, 1, 2$.}
\label{t:simple-blocks}
\end{table}

\subsection{Mapping $SU(N)$ models into F-theory}

We now return to the classification of $SU(N)$ building blocks for
anomaly-cancelling 6D chiral supergravity theories.  Associated with
each factor $G_i = SU(N_i)$ in the gauge group we have a set of matter
fields in representations satisfying (\ref{eq:f4-condition}),
(\ref{eq:a-condition}), (\ref{eq:c-condition}); these conditions
uniquely determine the coefficients $\alpha_i,\talpha_i$ for each
factor.  In Section \ref{sec:classification}, we showed how these
building blocks could be combined to construct complete lists of
anomaly-free models with multiple gauge group factors.  In this
section, we show how the data from anomaly cancellation in the
low-energy theory, namely the $\alpha_i, \talpha_i$, determine the
structure of the F-theory compactification.

In order to define the dictionary between the low-energy physics and
F-theory, we wish to associate with each gauge block a divisor $\xi_i$
on an appropriate Hirzebruch surface, to be used as the base of the
elliptic fibration in F-theory.  A block specified by $\alpha,
\talpha$ is mapped to the divisor
\begin{equation}
(\alpha,\talpha) \; \rightarrow
\;\xi =\frac{\alpha}{ 2}  (D_v + \frac{m}{2} D_s)
+ \frac{\talpha}{ 2}  D_s \,.
\label{eq:map}
\end{equation}
For example, an $SU(5)$ gauge group with 3 matter fields in the
antisymmetric tensor representation has $\alpha, \tilde{\alpha} = 2,
1$, and can be mapped to the divisor 
\begin{equation}
\{SU(5), 3 \;{\tiny\yng(1,1)}+ 19 \;{\tiny\yng(1)},
\alpha = 2, \tilde{\alpha} = 1\} \;
\rightarrow
\; (D_v +\frac{m}{2} D_s) +\frac{1}{2} D_s,
= D_v + \frac{m+1}{2}D_s
% % \label{eq:}
\end{equation}
This divisor corresponds to the class of an irreducible curve only for $m=1$.  In this case, the block specifies both the divisor $\xi$ and the base $\F_m$.

The correspondence between divisors in F-theory and the coefficients
$\alpha, \tilde{\alpha}$ was expressed in related forms in
\cite{Morrison-Vafa, Bershadsky-Vafa, Grassi-Morrison}.
To check that the map defined by (\ref{eq:map}) is correct, we compare
with the results of \cite{Grassi-Morrison}, from which we have
\begin{eqnarray} \label{eq:xiK}
\xi_i \cdot (-K) & = &  \frac{1}{6} \left( \sum_R x^i_{R} A_{R}^i -
A_{Adj}^i\right) = \alpha_i+\talpha_i\\
2\xi_i \cdot \xi_i & = & \frac{2}{3}\left( \sum_R x^i_{R} C_{R}^i -
C_{Adj}^i \right) =\alpha_i\talpha_i \,.
\label{eq:xixi}
\end{eqnarray}
A short computation using (\ref{eq:map}), (\ref{eq:f-k}) confirms that
these equations are satisfied.  Furthermore, we can check that the
product of two distinct blocks satisfies
\begin{equation}
4 \xi_i \cdot \xi_j = \alpha_i\talpha_j +\alpha_j\talpha_i \,.
% % \label{eq:}
\end{equation}

The conditions in equations (\ref{eq:xiK}) and (\ref{eq:xixi})
immediately guarantee that $\alpha_i$ and $\talpha_i$ are real.
This follows from 
the discriminant of the corresponding quadratic equation
\begin{equation}
(\xi_i \cdot K)^2 - 4\cdot 2\xi_i\cdot \xi_i
\end{equation}
being non-negative.  But that discriminant is the negative of the 
determinant of the matrix
\begin{equation} \label{eq:HImatrix}
\begin{bmatrix} \xi_i\cdot \xi_i\ & \ \xi_i\cdot K\\
\xi_i\cdot K \ & \ K\cdot K
\end{bmatrix}
\end{equation}
(since $K\cdot K=8 $ for a Hirzebruch surface $\mathbb F_m$), and
the Hodge index theorem for algebraic surfaces (cf.~\cite{BPV})
implies that (\ref{eq:HImatrix}) has negative or zero determinant.

Equations (\ref{eq:xiK}) and (\ref{eq:xixi}) are also the key to
generalizing the analysis in this paper to cases with $n_t>1$,
something we plan to pursue in future work.

Note that the map (\ref{eq:map}) can generally take a block to several
different choices of $\F_m$.  Furthermore, for some blocks, both
choices of $\alpha, \tilde{\alpha}$ lead to acceptable divisors.  For
example, an $SU(N)$ group with 2 antisymmetric representations has
either $\alpha = 2, \tilde{\alpha} = 0$ corresponding to the divisor
$D_v$ on $\F_0$ or $\alpha = 0, \tilde{\alpha} = 2$ corresponding to
the divisor $D_s$ on any $\F_m$.  For single-block models of this
type, there are distinct realizations on $\F_0, \F_1,$ and $\F_2$.  In
some cases apparently distinct realizations of a given model are
actually equivalent by a duality.  For example, $D_v$ and $D_s$ on
$\F_0$ are related by exchanging the two $\Pp^1$'s whose product forms
$\F_0$, corresponding to S-duality of the supergravity theory. In addition, 
$\F_2$ can be deformed to $\F_0$ through a complex structure deformation as 
discussed in \cite{Morrison-Vafa}, so models on these two surfaces may be 
related by deformations on a single moduli space.  It would be good to have a
general understanding of when distinct embeddings of a given model are
physically equivalent under a duality symmetry and when they are not.

In Table~\ref{t:simple-blocks}, we list the $SU(N)$ blocks with
fundamental and antisymmetric representations allowed by anomaly
cancellation, and the corresponding F-theory divisors on $\F_0, \F_1,
\F_2$.  The reason we restrict to $m=0,1,2$ is that at a general point
in moduli space, the gauge group is completely broken for blocks
listed in Table~\ref{t:simple-blocks}.  For other values of $m$, the
F-theory compactification has a nonabelian unbroken symmetry of a type other than $SU(N\geq 4)$ at a
general point in moduli space.  The table includes all blocks which
have $\nhv \leq 244$; larger values of $a$ are possible in models with
multiple gauge group factors.

Given the map on building blocks associated with gauge group factors,
in principle we can build up an arbitrary model with any product gauge
group and matter content from knowledge of the embedding of the blocks.
For example, the F-theory construction of the model with gauge group
$SU(4) \times SU(5)$ with 2 and 4 antisymmetric representations of the
factors described in (\ref{eq:product-example}) is associated with the
following set of singular divisors on the F-theory side, using the
base $\F_0$
\begin{eqnarray}
X_1 & = &4 \xi_1,  \hspace*{0.1in}  \xi_1 =   D_s\\
X_2 & = & 5 \xi_2,  \hspace*{0.1in}  \xi_2 =   D_v + D_s\\
Y & = & -24 K- 4 \xi_1 -5 \xi_2 = 19D_v + 15D_s \,.
\end{eqnarray}
Note that this same model could be constructed in two other ways.  The
same model can be realized on $\F_2$, where the map (\ref{eq:map})
gives $\xi_2 = D_v + 2D_s$.  Alternatively, we could have chosen
$\alpha_1 = 2, \tilde{\alpha} = 0$, giving the same gauge group and
matter content, but with the F-theory realization having $\xi_1 = D_v$
on $\F_0$.  As this example illustrates, some models have several
distinct realizations in F-theory.  A similar redundancy was noted in
\cite{Kumar-Taylor}, where multiple UV realizations of some specific
anomaly-free models were found in the heterotic string, associated
with topologically distinct lattice embeddings.  It would be nice to
have a better understanding of the physical differences between
different F-theory realizations of the models considered here.

As another example of how a complete model is mapped to F-theory using
(\ref{eq:map}) consider the model with gauge group $SU(16) \times
SU(4)^8$ described below equation (\ref{eq:16-48}).  The $SU(16)$ has
$a = 0$ and maps to (16 copies of) $D_v$ on $\F_2$.  Each $SU(4)$ 
has $a = 2$ and maps
to (4 $\times$) $D_s$.  The bifundamental in each $SU(16) \times SU(4)$ follows
from the intersection number $D_v \cdot D_s = 1$.  The total
singularity locus for this model is
\begin{equation}
\sum_i X_i = 16D_v + 32D_s \,.
% \label{eq:}
\end{equation}
Note that no more factors of $SU(4)$ can be added in F-theory because
then the residual singularity locus $Y = -12 K-\sum_{i} X_i$ could not
be expressed as a sum of irreducible components without further
singularities on $D_v$.

Given the map (\ref{eq:map}) we can compare the constraints on models
from anomaly cancellation to the geometric constructions on the
F-theory side.  It is remarkable how neatly specific properties of the
anomaly equations are mirrored in the F-theory geometry.  For example,
on the anomaly side, we know that it is not possible to have more than
one gauge group factor with a negative $\tilde{\alpha}$.  Thus, we
cannot have more than one $SU(N)$ factor with 0 or 1 antisymmetric
representations.  On the F-theory side, this corresponds to the fact
that the divisor $D_v$ on $\F_1$ and $\F_2$ has $D_v \cdot D_v = -m <
0$, associated with the fact that this divisor has no deformations.
Thus, all singularities associated with this topological equivalence
class are coincident, and only one $X_i$ of this type can appear in
the decomposition (\ref{eq:Kodaira-formula}).

The genus of an irreducible, non-singular curve in the class $\xi_i$ is
determined by the adjunction formula
\begin{equation}
K\cdot \xi_i + \xi_i\cdot \xi_i = 2g_i-2 .
\end{equation}
Note that by equations (\ref{eq:xiK}) and (\ref{eq:xixi}), this
can also be expressed as
\begin{equation}
g_i = \left( \frac12\alpha_i-1\right)\left(\frac12\talpha_i-1\right) .
\end{equation}
On the F-theory side, a genus $g$ curve corresponding to an $SU(N)$
gauge group gives $g$ hypermultiplets in the adjoint representation.
In the anomaly analysis of Section \ref{sec:classification} for
$SU(N)$ blocks with fundamental and 2-index antisymmetric matter, one
of $\alpha$ or $\talpha$ is equal to 2.  This fact implies that the
genus of a non-singular, irreducible curve in the class $\xi$
corresponding to the map (\ref{eq:map}) is always zero.  This is in
agreement with the fact that there are no adjoint matter
hypermultiplets.  We give some examples of blocks with adjoint matter
in the following section.

Another property of the anomaly cancellation equations which is
mirrored neatly in the F-theory geometry is the fact that blocks with
values of $a$ differing in parity cannot appear in the same model,
except in special circumstances.  In particular, if one $SU(N)$ block 
has $(\alpha_1,\talpha_1) = (2, a_1-2)$ with $a_1$ odd, the
second group cannot have $(\alpha_2,\talpha_2) = (2, \mbox{even})$, or
the number of bifundamentals would not be integral. In F-theory, this parity 
constraint arises because for $a$ even/odd with $\alpha = 2$, the map
(\ref{eq:map}) gives a divisor on $F_m$ with $m$ even/odd. As a result, a second block of the above form $(2,\mbox{even})$ would map to a fractional divisor, which is not allowed. Thus, an $SU(N)$ block with $(\alpha_1, \talpha_1)=(2,a_1-2)$ where $a_1$ is odd, can be combined with another block with even $a_2$, only if $(\alpha_2,\talpha_2)=(a_2-2,2)$ with $a_2 \equiv 2 (\mbox{mod }4)$. For example, if $(\alpha_1,\talpha_1)=(2,\mbox{odd})$ and $(\alpha_2,\talpha_2)=(0,2)$, both blocks can be realized on $\F_1$, with the second block on $D_s$.

The map (\ref{eq:map}) defines a set of divisors in $\F_m$ for
any model with gauge group of the form ${\cal G} = \prod_{i} SU(N_i)$
and matter in the fundamental and antisymmetric representations.  In
the next section we discuss the extension of this embedding to other
representations and other groups.  First, however we discuss the
conditions which must be satisfied for the singularity locus defined
in this way to give the desired F-theory model.

To show that the models defined in F-theory as described above indeed
have the correct structure, we must first check that the matter
content of the theory is that desired.  Given a gauge group $SU(N)$
with $f$ fundamental and $a$ antisymmetric matter fields, we wish to
check that the F-theory model defined through the map (\ref{eq:map})
correctly reproduces these numbers of fields in each representation.
As shown in \cite{Grassi-Morrison}, indeed
\begin{eqnarray}
a & = &  \xi \cdot (-K) = \alpha + \tilde{\alpha}\\
f & = &  -8 \xi \cdot K-N \xi \cdot \xi
= 8a + N \alpha \tilde{\alpha}/2 = 8a + N (2-a)
\end{eqnarray}
in agreement with the $F^4$ relation (\ref{eq:ffa}).

For a complete model we must also show that $\nhv = 244$.  Because the
matter content of each gauge group is correctly reproduced by the
geometric model produced through (\ref{eq:map}), the only question is
whether the number of neutral hypermultiplets associated with the
residual discriminant locus  $Y = -12 K-\sum_{i} N_i \xi_i$ is the correct
number to saturate the gravity anomaly (\ref{eq:bound}).  Indeed, one
of the principal results of \cite{Grassi-Morrison} was the
demonstration that this gravitational anomaly is  precisely
saturated, based on an explicit calculation of the number of neutral
hypermultiplets arising from cusps in $Y$.

Thus, we have shown that for any model composed of simple blocks of
the type considered so far, the map (\ref{eq:map}) gives an
appropriate combination of divisor classes in $\F_m$.  From the
definition of the map, then, it seems plausible that we can construct
an F-theory model for any anomaly-free supergravity theory in the class considered so far.  To show
this conclusively, however, we must check several things.

\noindent
1) We must show that there are no consistent supergravity models such
   that the image in F-theory requires a sum of divisors so large that
   the residual discriminant locus $ Y$ cannot be written as a sum of
   effective irreducible divisors.  In such a situation, such as if
   $\sum_{i} X_i = aD_v + bD_s$ with $a > 24$, there would not be an
   F-theory description of the complete model.  We have checked that
   the map (\ref{eq:map}) leads to an acceptable set of divisors for
   all of the 16,418 $SU(N)$ models explicitly tabulated in
   Table~\ref{t:block-models}.  It would be nice to have a more
   general proof that this always works.

\noindent
2) Even if all consistent supergravity models lead to configurations
   with acceptable $Y$'s, we have only described the topological
   structure of the singularity locus.  To guarantee that the model is
   well-defined, we need a Weierstrass model explicitly describing the
   elliptic fibrations (or some other equally explicit description).
   We believe that such a Weierstrass model should exist for any
   configuration of divisors satisfying the anomaly cancellation
   conditions (in particular (\ref{eq:bound})).  We return to this
   question in Section \ref{sec:Weierstrass}.

\section{More representations and groups}
\label{sec:more-representations}

While we defined the map (\ref{eq:map}) from supergravity building
blocks to F-theory divisors above in the context of $SU(N)$ blocks
with only fundamental and antisymmetric matter, it seems that (up to a
constant) this map immediately provides a correct embedding of most  6D
chiral supergravity models in F-theory.

In this section we expand the map to include more general $SU(N)$
 matter representations as well as other gauge groups.  We give
 examples of various other matter representations and gauge groups,
 and describe their embedding in F-theory.  This works in most cases,
 but there are some situations in which the image of a block in
 F-theory does not correspond to an integral divisor.  These models
 may not have F-theory representatives and may suffer from some kind
 of quantum inconsistency.  In other cases we find exotic matter
 representations for which no corresponding singularity structure has
 yet been identified in F-theory.  We do not attempt a comprehensive
 analysis here of all possible gauge group and matter blocks, but give
 examples which display the generality of the supergravity-F-theory
 map (\ref{eq:map}).

\subsection{Other representations of $SU(N)$}

\subsubsection{Adjoint representation}

As mentioned above, in F-theory the genus $g$ of the divisor
determines the number  $d$ of adjoint matter representations
transforming under the group associated with that divisor.  In 6D
supergravity, we can include $d$ adjoint matter representations for
the group $SU(N)$.
The adjoint of $SU(N)$ has $A = B = 2 N, C = 6, D = N^2 -1$.  Thus,
with the addition of $d$ adjoints the
relation (\ref{eq:ffa})
between the number of fundamentals and $N, a$ becomes
\begin{equation}
f =2 N -2 N d -a (N -8)  \,.
%% \label{eq:}
\end{equation}
The anomaly equations (\ref{eq:a-condition}) and
(\ref{eq:c-condition}) in the presence of adjoint matter are 
\begin{eqnarray}
\alpha+\talpha & = &  a\\
\alpha\talpha  & = &  2a + 4 (d-1)\,.
\end{eqnarray}
The solutions to these equations are
somewhat
limited.  
For example, for one adjoint ($d = 1$) we have 
$\alpha \tilde{\alpha} = 2a$.  In this case
the solutions for $\alpha, \tilde{\alpha}$ are only real when $a \geq
8$.  For $a = 8$ we have blocks associated with $SU(N)$ with $(\alpha,
\tilde{\alpha}) = (4, 4)$.  This maps using (\ref{eq:map}) to $2D_v +
3D_s$ on $\F_1$ which indeed is a genus 1 divisor.  Similarly, for $a
= 9$ we have $(\alpha, \tilde{\alpha}) = (6, 3)$ which maps to $3 (D_v
+ D_s)$ which is a genus 1 divisor on $\F_1$.

The story becomes more unusual for $a = 10$, where we have $f = -10 (N
-8)$.  If we choose $N = 4$, there is a single block model with $\nhv
= 220$, where $\alpha, \tilde{\alpha} = 5 \pm \sqrt{5}$.  Since these
$\alpha$'s are not rational, the map (\ref{eq:map}) does not take them
to divisors on any $\F_m$.  Thus, the one-block model with gauge group
$SU(4)$, one adjoint, 10 antisymmetric and 40 fundamental matter
hypermultiplets seems to satisfy anomaly cancellation and has gauge
kinetic terms with the correct sign but does not seem to have an
embedding in F-theory. We comment further on this and other models with irrational $(\alpha,\talpha)$ in Section \ref{sec:summary}.

We can perform a similar analysis for $d = 2$.  The smallest value of
$a$ for which $\alpha, \tilde{\alpha}$ are real is $a = 10$, for which
$(\alpha, \tilde{\alpha}) = (4, 6)$ (in either order).  This could
correspond to various divisors such as $2D_v + 3D_s$ on $\F_0$, all of
which have genus $g = 2$.

Because a single adjoint matter hypermultiplet has the same dimension
as the vector multiplet, the contribution to $\nhv$ from any block
with at least one adjoint matter multiplet is necessarily positive,
and is at least $N f \geq N f/2$.  Thus, the same algorithm as used in Section
\ref{sec:supergravity} can be used to classify and enumerate all
models including those with adjoint matter.

\subsubsection{3-index antisymmetric representation}

Now, consider including the 3-index antisymmetric representation,
which has (these constants, found in \cite{Erler}, can be reproduced
by simply considering the action of two orthogonal diagonal SU(N)
generators on the states labeled by Young tableaux).
\begin{eqnarray}
A_{3a} =\frac{1}{2} (N^2 -5 N + 6) &  &
B_{3a} =\frac{1}{2} (N^2 -17 N + 54)  \\
C_{3a} =(3 N -12) & \hspace*{0.3in} &  D_{3a} =\frac{1}{6} N (N -1) (N -2) \,.
\end{eqnarray}
Using these relations (\ref{eq:ffa}) is modified to
\begin{equation}
f =2 N -a (N -8) -\frac{1}{2}(N^2 -17 N + 54) \n3a \, ,
\label{eq:ffa3}
\end{equation}
where $t$ denotes the number of hypermultiplets in the 3-index antisymmetric representation. The anomaly polynomial again factorizes,  in the form
\begin{equation}
I = (\tr R^2 - 2 \tr F^2)(\tr R^2 -(a-2 + (N -4)\n3a)
\tr F^2) 
% \label{eq:}
\end{equation}
so (up to exchange) we have
\begin{equation}
\alpha = 2, \;\;\;\;\; \;\;\;\;\;
\tilde{\alpha} = a-2 + (N -4)\n3a \,.
%% \label{eq:}
\end{equation}

The contribution to the matter bound is
\begin{equation}
\nhv = 
 1 + N \left( f + a (N -1)/2 + t (N -1) (N -2)/6-N \right) \,.
\label{eq:mb3}
\end{equation}
Restricting to single blocks with $\nhv \leq 244$ there are solutions
for $N = 6, 7, 8$ (for $N < 6$ the 3-index antisymmetric representation is
equivalent to the fundamental, antisymmetric, or conjugate thereof).
For $N = 6$, there can be $\n3a = 1, 2$ or 3 fields in the
3-index antisymmetric representation, with $a$ up to $5, 3,$ or 1 in these
respective cases, for a total of 12 distinct models.  For $N = 7$ with
$\n3a = 1$ the range of $a$ is up to 3, and there is a model with
$\n3a = 2, a = 0$.  For $N = 8$ there are models with $\n3a = 1$ and
$a = 0, 1$.  Each of these models maps to a corresponding divisor in
F-theory.  For example, the $N = 8, a = 1$ model has $\tilde{\alpha} =
3$ so maps to $D_v + 2D_s$ in $\F_1$.  The singularity structure
corresponding to these matter representations for $N = 6, 7, 8$ is
described in F-theory in \cite{Grassi-Morrison-2}

If we extend to multiple-block models, there may be other possibilities.
For example, the block with $N = 9, \n3a = 1, a = 0$ has $f = 27$
fundamentals.  By itself, the contribution to $\nhv$ from this block
is 247, but it may be possible to combine this with other blocks in a
complete model.  The singularity type associated with a divisor of
this kind is unknown.  It would be interesting to either show that
this block cannot appear in a complete supergravity theory, or find an
F-theory realization of a model containing this block.

\subsubsection{Symmetric representation}
\label{sec:symmetric}

When we include $s$ symmetric representations the anomaly polynomial
no longer has an obvious algebraic factorization in general.
The $F^4$ anomaly condition is then modified from
(\ref{eq:ffa}) to
\begin{equation}
f =2 N -a (N -8) -s (N +8) \,.
\label{eq:ffas}
\end{equation}
Including symmetric representations as well as antisymmetric and
fundamental, a systematic analysis finds 44 single-block models with
various combinations of $f, a, N$.  One interesting set of cases is
when $a = 0, s = 1$. 
In this case, $f =N -8$  and the anomaly factorizes with
\begin{equation}
\alpha = 1, \;\;\;\;\; \;\;\;\;\; \tilde{\alpha} = -2 \,.
%% \label{eq:}
\end{equation}
In this case, the map (\ref{eq:map}) does not take the block to an
integral divisor on any $\F_m$.  On $\F_4$, the image is $D_v/2$.
This is 
another 
example of a block which does not have a clear
corresponding geometric structure in F-theory.  
Like the previous example 
it is characterized by its failure to give an integral divisor
under the map (\ref{eq:map}).

There are other configurations with symmetric representations which
are better behaved.  If we have $s \geq 1$ with $a > 8$, there
are a variety of solutions.  For example, for $ N = 4$ there are
one-block solutions with $s = 1, a = 9, \ldots,12$, as well as with
$(s, a) = (2, 13), (3, 15), (4, 17), (6, 20)$.  Other similar
solutions exist for $N$ up to 8.  As an example of a block of this
type we have
\begin{equation}
SU(4), s = 5, a = 18, f = 20, \alpha = 6, \tilde{\alpha} = 7
% \label{eq:}
\end{equation}
For these solutions the associated divisors are generally integral.

In F-theory, symmetric matter does not arise from a local enhancement
of the singularity and cannot be determined just from the topological
class of the singularity locus. 
When the curve of $A_{N-1}$ singularities is
itself singular with $s$ double points, we have $s$ symmetric
hypermultiplets \cite{Sadov}.  Since the map (\ref{eq:map}) only
determines the topological class of the discriminant locus, 
more information is needed to encode models with this type of matter
in F-theory.  This additional information about the
number of double points, must be included to explicitly construct a
Weierstrass model for a theory with matter transforming under the
symmetric representation of $SU(N)$.

\subsubsection{4-index antisymmetric representations}

We can consider still larger representations.  For example, it is
natural to consider the 4-index antisymmetric representation of $SU(N)$.
There are a couple of exotic blocks with $SU(8)$ gauge group and
matter content
\begin{eqnarray}
{\tiny\yng(1)}+ 3 \;{\tiny\yng( 2)}+ 2 \;{\tiny\yng(1,1,1)}
+  {\tiny\yng(1,1,1,1)} &  & (\nhv = 243) \\
{2\;\tiny\yng(1,1)}+ 3 \;{\tiny\yng( 2)}
+ 2 \;{\tiny\yng(1,1,1,1)} &  & (\nhv = 241)  \,.
\end{eqnarray}
Both these blocks have $(\alpha,\talpha)=(6,5)$. 
We are not aware of a singularity structure in F-theory which would produce the
4-index antisymmetric tensor representation, but it is 
%quite 
possible
that such an exotic singularity structure could exist.

\subsubsection{Larger representations}

As the matter representations become larger, the contribution to
$\nhv$ from these hypermultiplets increases. As a consequence, for
more complicated representations than those considered above there are
very few values of $N$ which do not immediately oversaturate the
$\nhv=244$ bound.  We have not attempted to completely classify the
supergravity blocks which may include these larger representations. We
leave the investigation of these more exotic models to future work.

\subsection{$SU(2)$ and $SU(3)$}
\label{sec:23}

Blocks with gauge group $SU(2)$, $SU(3)$ are special in the $SU(N)$
series, as they do not have an irreducible fourth-order invariant. In
addition, since $\pi_6(SU(2))=\Z_{12}$ and $\pi_6(SU(3))=\Z_6$, we
have to consider possible global anomalies
\cite{witten-global-anomaly}. We consider blocks with $SU(2)$ or $SU(3)$
gauge group and $f$ hypermultiplets in the fundamental
representation\footnote{The 2-index antisymmetric of $SU(2)$ is
trivial, and of $SU(3)$ is just the anti-fundamental}. These groups were
analyzed in \cite{Bershadsky-Vafa}, and we simply state the
results of applying the map (\ref{eq:map}) in these cases.  From the anomaly polynomial, the values of $\alpha,
\talpha$ are
\begin{align}
SU(2) :& \quad (\alpha,\talpha)=(2,\frac{f-16}{6}) \\
SU(3) :& \quad (\alpha,\talpha)=(2,\frac{f-18}{6})
\end{align}
Fractional values of $\alpha, \talpha$ map to non-integral divisor
classes under the map (\ref{eq:map}). 
This appears at first to give another class of non-integral
exceptional cases for the map to F-theory, but in this case
global anomalies constrain the
number of fundamental hypermultiplets modulo 6  through
\begin{align}
SU(2) :& \quad f \equiv \, 4\, (\rm{mod} \ 6) \nonumber \\
SU(3) :& \quad f \equiv \, 0\, (\rm{mod} \ 6) \label{eq:global-anomaly}
\end{align}
Thus,
the absence of global anomalies implies the integrality of $\alpha, \talpha$. 

The constraints from global anomalies in equation (\ref{eq:global-anomaly}), first derived in \cite{Bershadsky-Vafa}, can be understood from Higgsing. Consider a model with gauge group $SU(N)$ with $f$ fundamental and $a$ antisymmetric hypermultiplets. We can Higgs the gauge group down to $SU(N-1)$ by turning on a VEV for the fundamental 
hypermultiplets. Thus, we end up with a model with gauge group $SU(N-1)$ and $f'$ fundamentals and $a'$ antisymmetrics. The Higgsing can be worked out in the more familiar 4D, $\N=2$ language, and it turns out that $f'=f-2+a$, $a'=a$. Note that $f'=2(N-1) - a'((N-1)-8)$, which implies that Higgsing preserves the form of the $\tr F^4$ condition for $N \geq 4$. However, if we Higgs from $SU(4)\rightarrow SU(3)$, there is no $\tr F^4$ condition. Moreover, the 
antisymmetric representation is equivalent to the (anti) fundamental. Therefore, for $SU(3)$, $f'=f-2+2a \Rightarrow f'=6(a+1)$. This is in agreement with (\ref{eq:global-anomaly}). When the $SU(3)$ is then Higgsed down to $SU(2)$, we must have $f''=f'-2=4+6a$, which again agrees with (\ref{eq:global-anomaly}).

The gravitational anomaly requires that $f \leq 118$ for $SU(2)$ and $f \leq 84$ for $SU(3)$. We now check the validity of the divisor map (\ref{eq:map})  for the $SU(2)$ model with 118 fundamental hypermultiplets. 
\begin{equation}
(\alpha,\talpha) =(2,17) \longrightarrow D_v+9 D_s \mbox{ on } \F_1
\end{equation}
This does not oversaturate the Kodaira formula (\ref{eq:Kodaira-formula-2}). For the $SU(3)$ model with 84 fundamentals, 
\begin{equation}
(\alpha,\talpha) =(2,11) \longrightarrow D_v+6 D_s \mbox{ on } \F_1
\end{equation}
The divisor map (\ref{eq:map}) thus
works without exception for this class of $SU(2)$ and $SU(3)$ blocks.

We have incorporated all possible blocks with $SU(3)$ gauge groups
into the systematic analysis described in \ref{sec:classification}.
Including $SU(3)$ blocks increases the total number of possible models
to
68,997, with the number of models for a fixed number of factors
maximized at 20,639 models with 4 factors.  The largest number of
factors possible including $SU(3)$ blocks is 13, which occurs for a
single model with gauge group
\begin{equation}
G = SU(18) \times SU(3)^{12},
% \label{eq:}
\end{equation}
where there is a single bifundamental representation $(18, 3)$ for
each factor $SU(3)$, and no other matter fields transforming under any
of the gauge group components.  The F-theory map takes the $SU(18)$ to
$D_v$ on $\F_2$, and each $SU(3)$ factor to $D_s$. We discuss this case in more detail in Section \ref{sec:SU18-weierstrass}.
Note that the total of  68,997 models including $SU(3)$ blocks includes
46 models containing an SU(3) with no fundamental
matter.  Such a block has $(\alpha, \tilde{\alpha}) = (2, -3)$, and is
associated with the divisor $D_v$ on $\F_3$; the 46 models containing
this block can only be realized on $\F_3$.

\subsection{Tri-fundamental representation of $SU(M)\times SU(N)\times SU(P)$}

It is possible to have matter charged simultaneously under three
factors of the gauge group. The anomaly conditions constrain the
number of hypermultiplets that are simultaneously charged under two
factors of the gauge group. A tri-fundamental representation can occur
only if the anomaly conditions allow for sufficiently many
bifundamentals between every pair of groups. 
Through a complete enumeration of three block models with gauge groups $SU(M)
\times SU(N) \times SU(P)$, we find 848 models with one
hypermultiplet in the tri-fundamental representation.

In F-theory, a tri-fundamental of $SU(2)\times SU(2) \times SU(N)$ can
be realized if the three singular loci corresponding to the three
factors intersect at a point, and at that point the singularity type
is enhanced to $D_{N+2}$ \cite{Katz-Vafa}. 
Similarly, for $SU(2)\times SU(3)\times
SU(5)$, we would require the locus of $A_1$, $A_2$ and $A_4$
singularities to intersect at a point, with enhancement to $E_8$. In
an analogous manner, we can realize tri-fundamentals of $SU(2)\times
SU(3)\times SU(4)$ and $SU(2)\times SU(3)\times SU(3)$ through
enhancements to $E_7$ and $E_6$ respectively. In our exhaustive enumeration of three-stack models, we find that there are two models with gauge group $SU(2)\times SU(3)\times SU(6)$ with matter content
\begin{center}
$40 ({\tiny \yng(1)}, 1, 1) + 36(1, {\tiny \yng(1)}, 1) + 8(1,1,{\tiny \yng(1)}) +1(1,1,{\tiny \yng(1,1)})+ 1({\tiny \yng(1)},{\tiny \yng(1)},{\tiny \yng(1)})$  \\
$43 ({\tiny \yng(1)}, 1, 1) + 40(1, {\tiny \yng(1)}, 1) + 1({\tiny \yng(1)}, {\tiny \yng(1)}, 1)+ 1({\tiny \yng(1)},{\tiny \yng(1)},{\tiny \yng(1)})$.
\end{center}
The other models can all be realized using the singularity types
discussed above. We are not aware of the singularity structure in
F-theory that can realize the $SU(2)\times SU(3)\times SU(6)$
models. We postpone the analysis of these cases to future work.  One
possible realization of tri-fundamental matter fields might be through
string junctions (see, {\it e.g.}, \cite{DeWolfe-Zwiebach}) which end
on three 7-brane stacks and hence carry charge under three groups.

\subsection{SO(N)}
\label{sec:so}

So far, we have used the map (\ref{eq:map}) to take blocks with
$SU(N)$ gauge group to F-theory divisors.  In fact, essentially the
same map works for all simple groups, up to an overall constant which
depends upon the group.  In this section, we consider the case of $SO(N)$.  If we only have
fundamental representations (or bifundamental), then the $F^4$
condition gives
\begin{equation}
f = N - 8
% \label{eq:}
\end{equation}
and the anomaly polynomial factorizes as
\begin{equation}
I = (\rho -\phi) (\rho + 2 \phi) \,.
% \label{eq:}
\end{equation}
Thus, $\alpha, \tilde{\alpha} = 1, -2$ (in either order),
and the gauge group can only have one
such factor.
For a single SO(N) block we have $f N = N (N -8) \leq 244 + N (N -1)/2$
so $N \leq 30$.

%It seems that the normalization factor is different for different
%groups.  So we probably must take the map to give
For the gauge group $SO(N)$, we use the map 
\begin{equation}
\alpha (D_v + m/2 D_s) + \tilde{\alpha} D_s \,.
\label{eq:SO-map}
\end{equation}
Note that the normalization factor is different from the divisor map
(\ref{eq:map}) for the $SU(N)$ case by a factor of 2.  This
normalization factor depends on the choice of trace convention in the fundamental
representation.  We choose the normalization factor here to give an
integral divisor.  The divisor is irreducible and effective only for
$m=4$. F-theory on a CY 3-fold with base $F_4$ is dual to the $SO(32)$
heterotic string. At a general point in its moduli space, there is an
unbroken $SO(8)$ gauge group. This corresponds to the model with 0
fundamentals. The maximal gauge group $SO(30)$ can be realized with
the $SO(32)$ heterotic string, by choosing a $U(1)$ gauge bundle of
instanton number 24 \cite{gsw, Kumar-Taylor}. 
%This is an example of
%a family of anomaly-free models, {\it all} of which have a
%UV-completion as a string compactification.

In \cite{universality}, we found a model with gauge group
$SU(24) \times SO(8)$ with 3 hypermultiplets in the $({\tiny
\yng(1,1)}, 1)$ representation. The values of $(\alpha, \talpha)$ are
$(1,2)$ for the $SU(24) $ and $(1,-2)$ for the $SO(8)$. From the
divisor map (\ref{eq:SO-map}), the $SO(8)$ is realized on $D_v$ in the
base $F_4$. The $SU(24)$ singularity, however, is mapped to a
fractional divisor $\frac{1}{2} D_u$. 
This gives another example of an apparently anomaly-free supergravity
model with a block which maps to a non-integral divisor in F-theory.

\subsection{Exceptional groups}

For the $E_n$ groups, we can again
compute the map in the same way.  
From \cite{Erler} we have the following anomaly coefficients for the
fundamental and adjoint representations of $E_6, E_7, E_8$
\begin{center}
\begin{tabular}{|c|l|c | c | c | c |}
\hline
Group & Representation & $A_R$ & $B_R$ & $C_R$ \\
\hline
\multirow{2}{*}{$E_6$} & fundamental & 1 & 0 & $\frac{1}{12}$\\
 & adjoint & 4 & 0 & $\frac{1}{2}$  \\
\hline
\multirow{2}{*}{$E_7$} &fundamental & 1 & 0 & $\frac{1}{24}$\\
 & adjoint & 3 & 0 & $\frac{1}{6}$  \\
\hline
\multirow{2}{*}{$E_8$} &fundamental & 1 & 0 & $\frac{1}{100}$\\
 & adjoint & 1 & 0 &  $\frac{1}{100} $ \\
\hline
\end{tabular}
\end{center}
(Note that the adjoint of $E_8$ is equivalent to the fundamental, up
to a constant.)

Again, defining the divisor map for each gauge group requires a choice
of normalization constant.  Choosing constant factors $3, 6, 30$ for
$E_6, E_7, E_8$ gives the only possible map from these groups without
matter to acceptable F-theory divisors
\begin{eqnarray}
E_6: &  & (\alpha, \tilde{\alpha}) = \left(\frac{1}{3}, -1\right) \rightarrow D_v 
\; {\rm on} \;\F_6\\
E_7: &  & (\alpha, \tilde{\alpha}) = \left(\frac{1}{6}, - \frac{2}{3}\right) \rightarrow D_v 
\; {\rm on} \;\F_8\\
E_8: &  & (\alpha, \tilde{\alpha}) = \left(\frac{1}{30}, -\frac{1}{5}\right) \rightarrow D_v 
\; {\rm on} \;\F_{12}
\end{eqnarray}
Note that essentially the same choice of constant was made in the analysis of \cite{Grassi-Morrison} to relate geometric structure to anomaly conditions.
In general, then, the map is defined as
\begin{align}
E_6: & \quad (\alpha,\talpha) \longrightarrow 3\left[\alpha(D_v+\frac{m}{2} D_s)+\talpha D_s \right] \\
E_7: & \quad (\alpha,\talpha) \longrightarrow 6\left[\alpha(D_v+\frac{m}{2} D_s)+\talpha D_s \right]\\
E_8: & \quad (\alpha,\talpha) \longrightarrow 30\left[\alpha(D_v+\frac{m}{2} D_s)+\talpha D_s\right]
\end{align}
For an $E_n$ block with $f$ fundamental matter fields, we thus have
\begin{eqnarray}
E_6 :  (\alpha, \tilde{\alpha}) = \left(\frac{1}{3}, \frac{f-6}{6}\right)  &\longrightarrow & D_v + \frac{m+f-6}{2} D_s\mbox{ on } \F_m \label{eq:E6-map} \\
E_7:   (\alpha, \tilde{\alpha}) = \left(\frac{1}{6}, \frac{f-4}{6}\right)& \longrightarrow & D_v + \frac{m+2f-8}{2} D_s\mbox{ on } \F_m \label{eq:E7-map}
\end{eqnarray}
We can confirm that this map works by considering the heterotic string on a K3 surface at the point with gauge symmetry $E_7\times E_8$. This is obtained by having all 24 instantons in a single $SU(2)\subset E_8$, which breaks $E_8$ down to the maximal subgroup $SU(2)\times E_7$. From the index theorem, the matter content can be worked out to be 10 hypermultiplets (or 20 half-hypermultiplets) in the fundamental of $E_7$. Since this model corresponds to a point on the branch of the heterotic string with instanton numbers $(24,0)$, the dual F-theory construction has base $\F_{12}$. 
This is in agreement with the divisor map --- the $E_8$ block with no charged matter is realized on $D_v$ and the $E_7$ block with 10 {\bf 56} hypermultiplets is realized on $D_u$. 
More generally, we could have instanton numbers $(12-k, 12+k)$ in $E_8\times E_8$, and put all the instantons in a single $SU(2)$ subgroup of each $E_8$ factor, resulting in the gauge group $E_7\times E_7$. The matter content computed by the index theorem gives $(8-k)/2$ $(56,1)$ and $(k+8)/2$ $(1,56)$. To obtain fermions of the right chirality to form hypermultiplets, we need $ k \leq 8$. When $k$ is odd, we end up with a half-hypermultiplet, which is allowed as the 56 of $E_7$ is pseudoreal. From the divisor map (\ref{eq:E7-map}), the first $E_7$ is realized on $D_v+\frac{m-k}{2} D_s$, which is irreducible only for $m=k$, thus fixing $m$. The second $E_7$ is realized on $D_v+mD_s=D_u$. There is no bifundamental matter, in agreement with $D_u\cdot D_v=0$. This verifies the consistency of this map with known heterotic constructions through the F-theory-heterotic duality\cite{F-theory, Morrison-Vafa}.

\subsection{Non-simply laced groups}

A similar analysis to the previous cases gives the map for the
non-simply laced groups.  For $F_4$ and $G_2$, which have no quartic
invariant, including $f$ matter fields in the fundamental
representation,
we have
\begin{eqnarray}
F_4:  (\alpha, \tilde{\alpha}) = \left(\frac{1}{3},
\frac{f-5}{6}\right)  &\longrightarrow & D_v + \frac{m+f-5}{2}
D_s\mbox{ on } \F_m \label{eq:f4-map} \\ 
G_2:   (\alpha, \tilde{\alpha}) = \left(1, \frac{f-10}{6}\right)&
\longrightarrow & D_v + \frac{3m+f-10}{6} D_s\mbox{ on } \F_m
\label{eq:g2-map} 
\end{eqnarray}
In the case of $G_2$, we see that $f \equiv 1$ \; (mod 3) is needed
for an integer divisor on some $\F_m$, in agreement with global
anomaly cancellation conditions.

For $Sp(N)$ with $f$ fundamentals, the story is similar to $SO(N)$.
Cancellation of the $F^4$ anomaly gives
\begin{equation}
f = 2 N + 8,
% \label{eq:}
\end{equation}
and the values of $\alpha, \tilde{\alpha}$ are 2, -1, associated with
$D_v$ on $\F_4$.

\section{Realizations in F-theory}
\label{sec:Weierstrass}

As discussed in Section \ref{sec:F-theory}, the map from supergravity
models to F-theory gives the topological data of the discriminant
locus and singularity structure needed for the corresponding F-theory
construction, but this does not immediately lead to an explicit construction of these
elliptic fibrations through something like a Weierstrass model.
\vspace*{0.05in}

\noindent
{\bf Conjecture:}
{\it Every combination of effective divisors $X_i$ and residual
  divisor $Y$ associated through (\ref{eq:map}) with a 6D supergravity theory
  satisfying the anomaly conditions,
  including the gravity bound (\ref{eq:bound}) associated with the
  Euler character of the total space of the elliptic fibration as
  described in \cite{Grassi-Morrison}, gives rise to an explicit
elliptic fibration through a Weierstrass model.}
\vspace*{0.05in}

We do not have a proof of this conjecture in general.  
In a number of cases we have considered explicitly, however, the contribution
of $n_h-n_v$ to the total gravitational anomaly for a supergravity
block can be identified directly with the number of degrees of freedom
in the Weierstrass model which are fixed in imposing the desired
singularity structure on the associated divisor.
This suggests that there is a generic sense in which this conjecture
should hold, since in any model the number of unfixed degrees of freedom
in the Weierstrass model should correspond to the number of neutral
hypermultiplets in the corresponding supergravity theory.
We give a concrete example of how this works for a specific class of
Weierstrass models below.

Extending the map defined in this paper to all possible building
blocks with arbitrary simple gauge groups and matter content, along
with a proof of this conjecture, would suffice to prove the ``string
universality'' conjecture \cite{universality} for chiral 6D
supergravity theories, to the extent that all configurations of gauge
groups and matter fields allowed in consistent models could be
embedded in F-theory.
Note that for general models including arbitrary matter types and
non-simply laced groups, the construction of appropriate Weierstrass
models must include all appropriate singularity types and monodromies
to realize the supergravity matter content and gauge group.

%To show that the conjecture just stated is perhaps plausible, we give
%an example of a Weierstrass model for a gauge group with a single
%$SU(N)$ factor and no antisymmetric matter fields.  We show that the
%bound on $N$ beyond which no Weierstrass model exists coincides with
%the bound $N \leq 15$ beyond which the gravitational anomaly cannot be
%satisfied.
To demonstrate the plausibility of the above conjecture, we now give
some explicit examples of elliptic fibrations over $\F_m$
for single-block models with gauge group $SU(N)$.
We also consider some cases with gauge group
$E_6$ and $E_7$ with fundamental matter. We show that anomaly-free
supergravities in these classes can be realized as explicit F-theory
compactifications through Weierstrass models.

\subsection{Weierstrass Models on Hirzebruch surfaces}

We first review the basics of Hirzebruch surfaces as presented in
\cite{Morrison-Vafa, Bershadsky-all}.  The surface $\F_m$ is defined
as a $\P^1$ bundle over $\P^1$ as follows
\begin{equation}
\F_m:=\{(u,v,s,t)\in \C^4 \backslash Z: \ (u,v,s,t) \sim (\mu \lambda^m u, \mu v, \lambda s, \lambda t), \; \lambda, \mu \in \C^*\}
\end{equation}
$Z$ is the set of fixed points of the $\C^*$-action specified by $\lambda, \mu$.  The divisors $D_u$, $D_v$ and $D_s$ as discussed in Section \ref{sec:review}, correspond to the curves $u=0$, $v=0$ and $s=0$ respectively.  An elliptically fibered Calabi-Yau 3-fold on the base $\F_m$ can be specified by the Weierstrass equation
\begin{equation}
y^2 = x^3 + f(s,t,u,v) xz^4 + g(s,t,u,v)z^6 \label{eq:weierstrass}
\end{equation}
in the weighted projective space $\P^{2,3,1}$.  The functions $f,g$ are sections of the line bundles $-4K$ and $-6K$ respectively, where $K$ is the canonical bundle.  In this section, we consider fibrations over $\F_0, \F_1, \F_2$, where the fiber suffers an $A_{N-1}$ (type $I_{N-1}$) degeneration on the locus $v=0$.  In the coordinate patch $w=v/u, \ z=s/t$, the defining polynomials $f(w,z)$ and $g(w,z)$ take the form
\begin{align}
f(w,z) & = \sum_{i=0}^8 w^i f_{8-4m+m i}(z) \\
g(w,z) & = \sum_{j=0}^{12} w^j g_{12-6m+mi}(z)
\end{align}
The limits in the summations above need to be adjusted to ensure that all polynomials have non-negative degree. 

The degeneration locus of the elliptic fibration is given by the vanishing of the discriminant of the defining equation (\ref{eq:weierstrass}).
\begin{equation}
\Delta (w,z) = 4 f(w,z)^3 + 27 g(w,z)^2
\end{equation}
For the total space of the elliptic fibration to be Calabi-Yau, we
need $m \leq 12$. The number of degrees of freedom in $f, g$
associated with the
coefficients of the polynomials is shown in Table \ref{t:dof-Fm}. We have
subtracted the deformations that correspond to symmetries of
$\F_m$, and the overall scale in the discriminant. The dimension of
the automorphism group of 
$\F_0$ is 6, and that of
$\F_m$ for $m > 0$
can be computed to be $m+5$ using the
formula in \cite{Cox-Katz}. 
The number of hypermultiplets is generally one more than the number of degrees
of freedom in the Weierstrass polynomials since there is one
additional degree of freedom from the overall K\"ahler modulus of the
F-theory base, except for $F_2$ where one additional hypermultiplet is
missing when there is no gauge group on $D_v$.
In the specific case of $\F_2$, we show
by example how the neutral hypermultiplets from the supergravity  theory
exactly match
with the degrees of freedom available in the Weierstrass model.

\begin{table}
\centering
\begin{tabular}{|c|c|c|c|c|c|c|c|c|c|c|c|c|c|}
\hline
$m$ & 0 & 1 & 2 & 3 & 4 & 5 & 6 & 7 & 8 & 9 & 10 & 11 & 12 \\
\hline
%DOF & 244& 243& 242& 251& 268& 294& 318& 348& 376& 404& 433& 453& 482 \\
DOF & 243& 243& 242& 251& 271& 295& 321& 348& 376& 404& 433& 462& 491 \\
\hline
\end{tabular}
\caption{Degrees of freedom (DOF) in 
%terms of 
coefficients of polynomials that appear in the Weierstrass equation describing an elliptic fibration over $\F_m$.} \label{t:dof-Fm}
\end{table}

\subsection{$SU(N)$}

In order to have an $A_{N-1}$ degeneration on the locus $w=0$, we require that
 $\ord_{w=0}(\Delta)=N$ and $\ord_{w=0}(f)=\ord_{w=0}(g)=0$. In addition, Tate's algorithm \cite{Bershadsky-all} requires an auxiliary polynomial to factorize, corresponding to the $I_N$ split condition; we discuss this condition in the following section.  
If the discriminant is of the form $\Delta=w^N (p(z)+w q(z,w))$, the
 locus $w=0$ is intersected by the other component $p(z)+wq(z,w)=0$ at
 the zeroes of the polynomial $p(z)$.  At these points $z=\zeta$, the
 singularity type of the fiber is enhanced to $A_N$.  In terms of the low-energy theory, this implies that a matter hypermultiplet in the fundamental representation of $SU(N)$ is localized at every zero $\zeta$.  For 2-index antisymmetric matter, we require that the singularity type be enhanced to 
$D_N$ at special points on the locus $w=0$.  (At these points, $f$ and $g$
will vanish, whereas they do not vanish when the fiber is enhanced to $A_N$.)
In this section, we construct Weierstrass models on bases $\F_1$ and $\F_2$ with $A_{N-1}$ locus $w=0$, which correspond to models with gauge group $SU(N)$ and matter hypermultiplets in the fundamental and 2-index anti-symmetric representation (see Table \ref{t:simple-blocks}). At a general point in moduli space for these models, i.e. with a general choice of polynomials $f(w,z), g(w,z)$, the 
gauge group is completely broken.

\subsubsection{$\F_2$}

On $\F_2$, as shown in Table~\ref{t:dof-Fm}, the coefficients in the
polynomials $f, g$ encode 242 independent degrees of freedom.  With an
$A_{N-1}$ singularity along the locus $D_v \ (w=0)$ of $\F_2$, we can
realize models with gauge group $SU(N)$ and $N_f=2N$ hypermultiplets
in the fundamental representation.  The gravitational anomaly
condition requires that $\nhv=244$ (including neutral hypermultiplets), and this implies that $N\leq 15$.
The matter content requires that the discriminant take the form
\begin{equation}
\Delta(w,z) = w^N(p_{2N}(z) + wq(z)+\ldots), \quad N \leq 15
\end{equation}
where $p_{2N}(z)$ is a polynomial with $2N$ distinct zeroes $\zeta_i$ and $q(\zeta_i)\neq 0$.  This requirement in fact, uniquely picks out the base $\F_2$.

The functions $f(w,z), g(w,z)$ for the base $\F_2$ can be written as 
\begin{eqnarray}
f(w,z) & = & \sum_{i=0}^8 w^if_{2i}(z) \\
g(w,z) & = & \sum_{j=0}^{12} w^{j}g_{2j}(z) 
\end{eqnarray}
The discriminant is
\begin{eqnarray}
\Delta(w,z) & = & 4f^3 + 27 g^2 \\
& = & 4f_0^3 + 27 g_0^2 + w(12 f_0^2f_2+54g_0g_2)+\ldots  \equiv  \sum_{k=0}^{24} C_{2k}(z) w^k
\end{eqnarray}
The coefficients $C_{2k}(z)$ in the expansion of $\Delta$ are polynomials of degree $2k$ in $z$.  In order to have an $A_{N-1}$ singularity along $w=0$, we need to tune the polynomials $f_{2i}$ and $g_{2j}$ so that $C_0=C_2=\ldots=C_{2N-2}=0$.  With the first $N$ coefficients set to zero, the discriminant is of the form
\begin{equation}
\Delta = w^N( C_{2N}(z) + w C_{2N+2}(z)+\ldots)
\end{equation}
At the zeroes of $C_{2N}(z)$, the singularity type is enhanced from $A_{N-1}$ to $A_N$, and as discussed earlier, this leads to $2N$ matter hypermultiplets in the fundamental representation.  
(Note that since $-K\cdot D_v=0$, neither $f$ nor $g$ will vanish along
$D_v$, so there is no anti-symmetric matter.)

We can see how the discriminant can be made to vanish with 
$\ord_{w=0}(\Delta)=N$  order by order as follows.  
Since the overall scale in the discriminant polynomial does not matter, we can set $f_0=-3$ without using up any degrees of freedom. We can now fix $g_0=2$, without using any degrees of freedom as this just fixes the location of singularity at $w=0$. By choosing $g_2(z)  \equiv -f_2(z)$, we use up 3 
degrees of freedom and the discriminant is of the form
\begin{equation}
\Delta(w,z) = -9(f_2^2 - 12 (f_4 + g_4)) w^2 + \O(w^3)
\end{equation}
Next, by choosing
\begin{equation}
g_4(z) \equiv \frac{1}{12} \left(f_2^2 - 12 f_4\right)
\end{equation}
the discriminant can be made to vanish to order three, and we have used up another 5 
degrees of freedom. 
In this manner, by an appropriate choice of polynomial $g_{2k}(z)$, the coefficient $C_{2k}(z)$ in the discriminant can be made to vanish for $ k \leq 12$. Thus, in order to obtain a gauge symmetry $SU(N), \ N \leq 13$, we need to fix the polynomials $g_2, g_4, \ldots g_{2N-2}$ and therefore use up $N^2-1$ 
degrees of freedom. 
The number of residual degrees of freedom works out to
$242-N^2+1=243-N^2$, and these should correspond to neutral
hypermultiplets. This agrees beautifully with a similar calculation
from the anomaly:
including only charged hypermultiplets we have $n_h-n_v = N^2+1$, and we need to add $243-N^2$
neutral hypermultiplets to satisfy the gravitational anomaly
condition.  
Thus, we see that at each value of $N$, the number of neutral
hypermultiplets on the supergravity side precisely corresponds to
the number of unfixed degrees of freedom in the F-theory polynomials.  We expect
that this will be the case quite generally, so that a correspondence
can be made between the contribution of any supergravity block to
$n_h-n_v$ and the additional coefficients which must be fixed in the
Weierstrass polynomials to encode the corresponding singularity.  We
leave a general proof of this assertion as a challenge for the future.

The gravitational anomaly condition imposes $N \leq 15$. In the analysis above, we showed that $SU(13)$ gauge symmetry could be obtained by just using the $g_{2k}$ polynomials. In order to go further, we need to use the degrees of freedom in the $f_{2k}$ polynomials. The discriminant for $SU(13)$ gauge symmetry is of the form
\begin{equation}
\Delta(w,z) = C_{26} w^{13} + C_{28} w^{14} + C_{30} w^{15} + \ldots
\end{equation}
We have $243-13^2=74$ actual degrees of freedom in the polynomials
$f_{2k}$. It is easy to see that with an appropriate choice of
coefficients in these polynomials, $C_{26}$ and $C_{28}$ could
generically
be made to vanish, but not $C_{30}$. This agrees nicely with the computation from anomaly cancellation in the low-energy theory.
We have an explicit solution for the $SU(14)$ case, which we present below.
\begin{align}
f(w,z)= &-3 + 12 h_8 w^4 + 12 h_{10} w^5 + 6 h_{10} h_2 w^6 + 2 h_{14} w^7 + (-12 h_8^2 - h_{14} h_2) w^8 \\
g(w,z)= & \ 2 - 12 h_8 w^4 - 12 h_{10} w^5 - 6 h_{10} h_2 w^6 - 
 2 h_{14} w^7 + (24 h_8^2 + h_{14} h_2) w^8 \nonumber \\
 & + 24 h_{10} h_8 w^9 +  12 h_{10} (h_{10} + h_8 h_2) w^{10} + 
 4 (h_8 h_{14} + 3 h_{10}^2 h_2) w^{11} \nonumber \\
 &+ (-16 h_8^3 - 2 h_8 h_{14} h_2 + 
    h_{10} (4 h_{14} + 3 h_{10} h_2^2)) w^{12}
\end{align}
The $h_i$ are general polynomials in $z$. The discriminant is
\begin{align}
\Delta(w,z) = -36 w^{14} (h_{14}^2 - 3 h_{10} h_{14} h_2^2 + 3 h_{10}^2 (24 h_8 + f_6 h_2) + \O(w))
\end{align}
We have not found an explicit Weierstrass model for the $SU(15)$ case.
While this problem seems difficult algebraically, counting degrees of
freedom seems to indicate that this should be possible.  Furthermore,
while similar algebraic difficulties appear in an analogous
construction for $SO(N)$ theories on $\F_4$ at $N = 30$, we know that
the $SO(30)$ model has an explicit string construction, as mentioned
in Section \ref{sec:so}.  We leave the explicit construction of the
$SU(15)$ model in this family as a challenge for the future.

All the supergravity models in this family are related to the
(hypothetical) $SU(15)$ model by Higgsing; turning on a VEV for a
fundamental hypermultiplet in $SU(N)$, breaks the gauge group to
$SU(N-1)$ and a mass term is generated for two fundamental
hypermultiplets.  This shows that the number of $SU(N-1)$
hypermultiplets in the low-energy theory is $f-2=2(N-1)$, in agreement
with (\ref{eq:ffa}) for $a = 0$.

\subsubsection{$\F_1$}

In this subsection, we construct $SU(N)$ models with $N+8$ fundamental
hypermultiplets and one 2-index antisymmetric hypermultiplet. To
accomplish this, we engineer an $A_{N-1}$ singularity along the $w=0$
locus of $\F_1$, which corresponds to the divisor $D_v$ (see Table
\ref{t:simple-blocks}). 

The polynomials $f$, $g$ on $\F_1$ take the form --
\begin{align}
f(w,z) & = \sum_{i=0}^8 w^i f_{i+4}(z) \\
g(w,z) & = \sum_{j=0}^{12} w^j g_{j+6}(z)
\end{align}
The discriminant locus is of the form
\begin{align}
\Delta(w,z) = 4f_4^3+27 g_6^2 + w(12 f_4^2 f_5+54 g_6g_7)+\ldots \equiv
\sum_{k=0}^{24} C_{k+12}(z) w^k 
\end{align}
An $SU(N)$ singularity requires the discriminant to vanish at order $N$
on the locus $w=0$. We will see that once this singularity is
engineered, the matter content works out very nicely in accordance with
anomaly cancellation. To obtain $SU(4)$ gauge symmetry, we can choose
\begin{align}
\begin{array}{c}
f_4(z)=-3q_2(z)^2, \qquad  f_5(z) = p_3(z) q_2(z),   \\
g_6(z)=2q_2(z)^3,  \quad  g_7(z) = -f_5(z) q_2(z),  \quad g_8(z)=
\left(-f_6(z) +\frac{1}{12} p_3(z)^2 \right) q_2(z),  \\
g_9(z) = \frac{1}{216} \left(36 f_6(z) p_3(z) + p_3(z)^3 - 216 f_7(z)
q_2(z) \right) 
\end{array} \label{eq:SU4-case}
\end{align}
Here $q_2(z)$ and $p_3(z)$ are arbitrary polynomials of degree 2 and 3
respectively. For the singularity to produce $SU(4)$ gauge symmetry, the
polynomial $q_2(z)$ must be perfect square, so $q_2(z)=\lambda (z-z_0)^2
$. This corresponds to the split $I_4$ singularity in Tate's algorithm
discussed in \cite{Bershadsky-all}. With this choice, the discriminant
takes the form
\begin{equation}
\Delta(w,z)=w^4 \left[ (z-z_0)^4 C_{12}(z) + \O(w) \right]
\end{equation}
%(-144 f_6^2 - 24 f_6 p_3^2 - p_3^4 - 288 f_7 p_3 q_2 + 1728 q_2 (g_{10} + f_8 q_2)
Here $C_{12}(z)$ is a general polynomial of degree 12 with distinct
roots. The locus $w=0$ is intersected by the residual locus at the point
$z=z_0$. The singularity type is enhanced to $D_4$, and thus, we obtain
one antisymmetric tensor of $SU(4)$. 

In the general $SU(N)$ case, when $N=2k$ the structure of the singular
locus of the fibration is similar to the $SU(4)$ case. The discriminant
is of the form 
\begin{equation}
\Delta(w,z) = w^N \left[ (z-z_0)^4 C_{N+8}(z) + \O(w) \right]
\end{equation}
The fact that the polynomial $C_{N+8}(z)$ has $N+8$ distinct roots results in 
$N+8$ fundamental hypermultiplets, in agreement
with the anomaly calculation. At $z=z_0$, the singularity is
enhanced to $D_N$ (since $f$ and $g$ vanish there), 
and we have antisymmetric matter localized at this
point. When $N=2k+1$, however, the singularity structure is slightly
different, and we discuss this in the $SU(5)$ case. For an $SU(5)$
singularity, in addition to the choices made in (\ref{eq:SU4-case}), we
need
\begin{equation}
f_6 = -\frac{1}{12} p_3^2 + p_4 q_2, \qquad  g_{10}(z) = \frac{1}{12}
\left(2 f_7 p_3 - 12 f_8 q_2 + p_4^2 q_2\right)
\end{equation}
where $q_2(z)=\lambda (z-z_0)^2$, and $p_3,p_4$ are general polynomials
in $z$ of degree 3 and 4 respectively. The discriminant is
\begin{equation}
\Delta(w,z)=w^5 \left[ (z-z_0)^6 C_{11}(z) + \O(w) \right]
\end{equation}
At first sight, this appears to be at odds with the anomaly conditions,
since we would have only 11 fundamental hypermultiplets at the roots of
$C_{11}(z)$. It turns out that at the point $w=0,z=z_0$, the singularity
type is enhanced all the way to $D_6$ (split $I_2^*$ according to Tate's
algorithm). This enhancement results in an antisymmetric tensor of
$SU(5)$ and 2 fundamental hypermultiplets, so all together we still have
13 fundamental hypermultiplets as required by the anomaly conditions.
For the general $SU(N)$ case, when $N$ is odd, the discriminant takes
the form
\begin{equation}
\Delta(w,z) = w^N \left[ (z-z_0)^6 C_{N+6}(z) + \O(w) \right]
\end{equation}
At $z=z_0$ the singularity type is enhanced all the way to $D_{N+1}$ which provides
the additional 2 hypermultiplets in the fundamental.

The gravitational anomaly restricts $N \leq 15$, and as in the previous
case, this agrees with the counting of degrees of freedom in the
Weierstrass model. Again, we have explicitly constructed a Weierstrass model 
only for the $SU(14)$ case, though a count of the degrees of freedom suggest 
that an $SU(15)$ Weierstrass model should exist. 

The basic method of
construction followed here for singularities on the divisor $D_v$  of
$\F_{1, 2}$ is easily adapted for $A_{N -1}$ singularities on the base
$\F_m, \ m=0$ with divisor $D_v$ or on
$\F_m, \ m=0,1,2$ with divisor $D_u$ or $D_s$. In each case, we can find
Weierstrass models compatible with the topological data provided by the
map \ref{eq:map} for most values of $N$. 
Although in both the $\F_2$ and the $\F_1$ cases we encountered
algebraic difficulties in extending the construction to the maximum
value $N = 15$, in both cases a degree of freedom counting argument
suggests that solutions should exist.
Furthermore, as mentioned above the
existence of  a string construction for the analogous
$SO(30)$  model gives us additional confidence that the
Weierstrass models for $SU(15)$ blocks can be realized on $\F_1, \F_2$
despite the apparent complexity of the algebra in these cases.

\subsubsection{$SU(18)\times SU(3)^{12}$}
\label{sec:SU18-weierstrass}

In the systematic enumeration of models, including $SU(3)$ blocks,
described in Section \ref{sec:23}, the model with the greatest number of
blocks (13) has gauge group $SU(18)\times SU(3)^{12}$. The matter
content consists of bifundamental hypermultiplets charged under the
$SU(18)\times SU(3)$ for each $SU(3)$ factor. This model was first
constructed in a different context in \cite{braneCY}. The $SU(18)$
block contains a total of 36 fundamental hypermultiplets, all in
bifundamental representations, and thus belongs to the family in
Table \ref{t:simple-blocks} with $SU(N)$ gauge group and $2N$
fundamentals. In the case of single block models, the gravitational
anomaly $\nhv \leq 244$ restricted the gauge group to $SU(15)$ in this
family. In this case, however, the other factors contribute negatively
to $\nhv$ because the matter hypermultiplets are all ``shared''
between the various gauge factors.

From the map (\ref{eq:map}), as stated in \ref{sec:23},
we know that this model can be realized
on $\F_2$, with the $SU(18)$ factor on $D_v$ and the various $SU(3)$
factors on $D_s$. 
It is possible to construct a Weierstrass model for this combination
of singularities.
The polynomials $f$ and $g$ are given by
\begin{align}
f(w,z) & = -3 (h_0 +  h_2 w+ h_4 w^2) (9 (h_0 +  h_2 w+ h_4 w^2)^3-2 h_{12} w^6) \\
g(w,z) & =  54 (h_0 + h_2 w + h_4 w^2)^6 + h_{12}^2 w^{12} - 18 h_{12} w^6 (h_0 + h_2 w+ h_4 w^2)^3 
\end{align}
where $h_i$ are polynomials of degree  $i$ in $z$.
The discriminant is
\begin{equation}
\Delta(w,z) = -27 h_{12}^3 w^{18} (4 (h_0 + h_2 w+ h_4 w^2)^3
-h_{12} w^6 ) \,.
\end{equation}
It is clear that the $w=0$ locus gives an $SU(18)$ gauge symmetry. In
addition, at each zero of $h_{12}(z)$, we have an $SU(3)$ gauge
symmetry since the discriminant vanishes at third order. Each $SU(3)$
locus of the form $z = z_\alpha$, where $z_\alpha$ is a root of
$h_{12}(z)$, intersects the $w=0$ locus with the $SU(18)$ gauge
symmetry at one point. This gives one bifundamental between $SU(18)$
and each $SU(3)$ factor. 
The fact that even for the largest multi-block model, the map
(\ref{eq:map}) gives an acceptable set of F-theory divisors which
admit an explicit
Weierstrass model
provides contributing evidence for the
conjecture that all 
models with  topologically
acceptable divisors can be explicitly realized in F-theory.

\subsection{$E_6$}

For $E_6$ gauge symmetry on the locus $w=0$, we need $\ord(f) \geq 3$,
$\ord(g)=4$ and $\ord(\Delta)=8$. The locus $w=0$ is intersected by
other components of the discriminant locus, and at these points the
singularity type is enhanced to $E_7$. This implies that a fundamental hypermultiplet is localized at
every such intersection \cite{Bershadsky-all, Katz-Vafa}. In this section, we give explicit Weierstrass
models of $E_6$ gauge symmetry with fundamental matter. The divisor
map (\ref{eq:E6-map}) determines the divisor on $\F_m$, given the number
$f$ of fundamentals.
\begin{equation}
f \rightarrow D_v + \frac{m+f-6}{2} D_s
\end{equation} 
We focus on the case where the $E_6$ symmetry is realized on $D_v$ for simplicity. This is the case when $f=6-m$, $m=0,1,2,3,4$. In the neighborhood around $w=0$, 
\begin{align}
f(w,z) & = w^3 f_{8-m}(z)+\ldots+w^8 f_{8+4m}(z) \\
g(w,z) & = w^4 g_{12-2m}(z) + w^5 g_{12-m}(z)+\ldots+w^{12}g_{12+6m}(z)
\end{align}
The discriminant locus is of the form
\begin{equation}
\Delta(w,z) = w^8[27g_{12-2m}(z)^2+w(54g_{12-2m}(z)g_{12-m}(z)+4f_{8-m}(z)^3) + \ldots)]
\end{equation}
As explained in \cite{Bershadsky-all}, the polynomial $g_{12-2m}(z)=g_{6-m}(z)^2$ in order to obtain an $E_6$ singularity. The fundamentals of $E_6$ are localized at the zeroes of $g_{6-m}(z)$.

\subsection{$E_7$}
For $E_7$ gauge symmetry on the locus $w=0$, we need $\ord(f) = 3$, $\ord(g) \geq 5$ and $\ord(\Delta)=9$. For $f$ fundamental hypermultiplets, we need the $E_7$ singularity to enhance to $E_8$ at $f$ distinct points. From the divisor map (\ref{eq:E7-map}), an $E_7$ singularity on $D_v$ in $\F_m$ realizes models with $\frac{8-m}{2}$ fundamentals. 
\begin{align}
f(w,z) & = w^3 f_{8-m}(z)+\ldots+w^8 f_{8+4m}(z) \\
g(w,z) & = w^5 g_{12-m}(z)+\ldots+w^{12}g_{12+6m}(z)
\end{align}
The discriminant locus is of the form
\begin{equation}
\Delta(w,z) = w^9[f_{8-m}(z)^3+w(3f_{8-m}(z)^2f_{16}(z)+g_{12-m}(z)^2) + \ldots)]
\end{equation}
The adjoint representation of $E_8$ branches under the maximal subgroup $E_7\times SU(2)$ as
\begin{equation}
\mathbf{248}=(\mathbf{133},1)+(1,\mathbf{3})+(\mathbf{2},\mathbf{56})
\end{equation}
The $\mathbf{56}$ is pseudoreal, and so we have a half-hypermultiplet localized at the $8-m$ zeroes of $f_{8-m}(z)$, or equivalently $\frac{8-m}{2}$ hypermultiplets.

\section{Some exceptional cases}
\label{sec:summary}

In this paper we have found an explicit map from six-dimensional
chiral supergravity theories to topological data for F-theory
constructions.  This map seems to give a realization of a significant
fraction of the finite number of anomaly-free chiral 6D supergravity
theories with one tensor multiplet in terms of the F-theory limit of
string theory.

We have, however, encountered a number of exceptional cases in which
the map does not give a well-defined geometry in F-theory.  In this
section we briefly summarize some of the types of cases encountered.
This list is presumably not comprehensive, as we have only explored
some groups and representations.  It seems likely that there are a
number of other types of gauge groups and matter blocks which share
the features of these exceptional cases.  There may even be more
unusual classes of exceptions which we have not encountered.

Understanding whether these exceptional cases represent situations in
which there are quantum inconsistencies in apparently reasonable
classical low-energy models, or as-yet undiscovered types of string
compactifications, will hopefully be a productive way of extending our
understanding of the correspondence between string theory and
low-energy supergravity theories in six dimensions.

Some of the cases we have found in which the map from supergravity
blocks to topological F-theory data does not give well-defined
integral divisor classes are the following:

\begin{enumerate}

\item[{\it a})] For $SU(N)$ with $(N-8) \ {\tiny \yng(1)} + 1 {\tiny \yng(2)}$ matter hypermultiplets, the image divisor seems to have a
component $\frac{1}{2}D_v$.  

\item[{\it b})] Similarly, the $SU(24)$ block in the anomaly-free $SU(24) \times SO(8)$ model with $3 ({\tiny \yng(1,1)},1)$ hypermultiplets encountered in \cite{Kumar-Taylor} gives a divisor which is 1/2 of $D_u$.

\item[{\it c})] For $SU(N)$ with one adjoint and $10\ {\tiny \yng(1,1)} + 10(8-N)\ {\tiny \yng(1)}$, we get irrational values
$\alpha = 5 \pm \sqrt{5}$ for the $\alpha$'s, which do not map to a
  divisor with integer coefficients.

\end{enumerate}

The common thread in these exceptional cases is that the image of the
block through the map (\ref{eq:map}) is not an integral divisor in the
$\F_m$ base of the F-theory compactification.  We encountered one
class of cases in which such potential exceptions are already ruled
out by a known mechanism: for $SU(2)$ and $SU(3)$ with $f$
fundamentals, the image of the map is only an integral divisor if $f$
is congruent to 4 or 0 modulo 6.  In these cases, the blocks whose
images would correspond to non-integral divisors are ruled out by
global anomaly cancellation requirements.

It seems possible that other quantum consistency conditions may rule
out the low-energy theories associated with the other exceptional
cases listed above.
This may arise from some other kind of global anomaly or related
mechanism.  Or, since the terms in the action proportional to $\alpha$
have the flavor of Chern-Simons terms, it is possible that some
mechanism analogous to the quantization of Chern-Simons level may
enforce an integrality condition on the coefficients $\alpha,
\tilde{\alpha}$.  Such an argument certainly seems plausible in ruling
out the type of exceptional case exemplified in case  {\it c}, with irrational
values of $\alpha, \tilde{\alpha}$.  On the other hand, there may be
some other topological class of string theory compactifications, for
example in another discrete part of the moduli space (as considered in
\cite{tri}),  perhaps corresponding to a compactification on a space with
some discrete quotient structure, which gives rise to the models which
appear to have a half-integral divisor in the image of the map from
the supergravity blocks.  Understanding these exceptional cases better
should be a fruitful direction for future research.

In addition to these cases in which the image of the
supergravity block is not an integral divisor, we have encountered a
number of exotic representations whose F-theory geometry is not yet
understood.  For example, we found configurations with 4-index
antisymmetric representations of $SU(8)$ and others with
trifundamental representations of groups like $SU(2) \times SU(3)
\times SU(6)$, which do not correspond to any known geometric
structure in F-theory.  These also are interesting cases for future study.

\section{Conclusions}
\label{sec:conclusions}

In this paper we have described an explicit mapping from the set of
low-energy chiral six-dimensionful supergravity theories (with one
tensor multiplet and nonabelian gauge group) to F-theory.  This gives
a global picture of how low-energy theory and string theory are
connected in a reasonably tractable component of the string landscape.
Further study of this correspondence promises to shed light both on
the set of allowed string theory compactifications and on constraints
satisfied by low-energy supergravity theories with UV completions.

Following the proof \cite{finite} that there are only a finite number
of possible gauge groups and matter content for such theories, the
results presented here represent a further step towards proving the
conjecture stated in \cite{universality} that all UV-consistent 6D
chiral supergravity theories can be realized in string theory.  There
are a number of issues which must be clarified to make further
progress in this direction.  First, we have not systematically
enumerated all the possible 6D supergravity theories, and the gauge
group and matter types which can appear in such theories.  This can in
principle be done.  The enumeration of the finite set of possible
models on the supergravity side seems quite tractable computationally.
Second, given such an enumeration it would be necessary to identify
the structures in F-theory corresponding to all matter representations
appearing in the list.  We have identified in this paper a number of
matter representations whose corresponding geometry is not yet known;
the finite number of such exotic representations appearing in
acceptable 6D supergravity models should provide a good guide to
understanding the corresponding allowed singularity structures in
F-theory.  Third, we have found a number of situations where the image
of the map is not an integral divisor in F-theory.  Some of these are
summarized in the previous section.  Showing that these exceptional
cases are associated with quantum inconsistencies, or new string
vacua, would be necessary to complete the global picture of the map
described here.  Fourth, we have not shown that explicit Weierstrass
models are possible for all topological F-theory constructions, although we 
have shown this to be possible for certain families. In certain cases, we have a dimension-counting argument which supports the conjecture stated in
Section \ref{sec:Weierstrass} that all topologically allowed models in
the image of the map can be realized explicitly through Weierstrass
models. It would be nice to have a more general argument along these lines.  Finally, as mentioned above, we have restricted attention so
far to nonabelian models with one tensor multiplet; it is clearly of
interest to expand the analysis to include multiple tensor multiplets
and $U(1)$ factors in the gauge group.

There is an enormous literature on how different approaches to string
compactifications can give rise to different low-energy field theories
coupled to gravity in various dimensions, including the
six-dimensional case considered here and the four-dimensional case of
physical interest.  In particular, in \cite{Morrison-Vafa,
Bershadsky-all, Grassi-Morrison}, a detailed analysis was made of the
different singularity structures in F-theory and the associated gauge
structure and matter content in the associated low-energy theory.  In
most of this work the emphasis has been on going from string theory to
the low-energy theory.  In this paper, we have approached the problem
from the other direction, by formulating a map from the space of
low-energy theories to the space of string theories.  Both approaches
lead to valuable lessons about the connection between low-energy
theory and string theory.  It seems likely, however, that further
progress in understanding the map from low-energy theories to string
theory may be of particular value both in explicit model-building
efforts and in understanding the general structure of the landscape.

The map we have described in this paper from supergravity theories to
topological F-theory data is not unique in all cases.  For some
combinations of gauge group and matter content, there are different
ways of mapping the theory to F-theory, either by choosing distinct
base spaces $\F_m$, or by switching the values of $\alpha$ and
$\tilde{\alpha}$ in  the gauge group factors.  Thus, there may
be multiple F-theory models with given gauge group and matter content.
In some cases these F-theory models are related through a known
duality symmetry, but in other cases they are not.  In general, the
number of discrete choices for a given supergravity theory is fairly
small.  A similar phenomenon was found in \cite{Kumar-Taylor}, where
for many models the heterotic realization was uniquely determined by a
lattice embedding satisfying certain criteria, but in some cases
multiple distinct lattice embeddings give rise to distinct string
theory realizations of a specific gauge group and matter content.  We
have not explored in detail how these models or the distinct F-theory
realizations found here would differ, or when in general such
models are related by a duality symmetry; we leave exploration of these questions for future work.  Note that in principle it is possible
to imagine many distinct low-energy Lagrangians for theories with the
same gauge group and matter content, but more detailed considerations
may place constraints on which Lagrangians lead to consistent
theories.  We have not explored this issue here either, having focused
essentially only on the topological data of the models studied here.

In this paper we have focused on chiral six-dimensional supergravity
theories, which are strongly constrained by anomaly cancellation.  In
developing a dictionary connecting the low-energy supergravity
theories to string theory, we find explicit relationships between the
constraints imposed by the framework of string compactifications and
the anomaly cancellation constraints in 6D.  In other situations, such
as non-chiral six-dimensional supergravity theories, or general
supergravity theories in four dimensions, there are no gravitational
anomalies, and the constraints we know of on low-energy theories are
weaker.  Nonetheless, in these cases there are similar constraints on
the space of string compactifications.  By understanding the
dictionary between low-energy theories and string theory more clearly
in the chiral six-dimensional case, it may be possible to generalize
this dictionary to other cases in which the low-energy constraints are
less well understood.  In particular, for ${\cal N} = 2$ non-chiral
supergravity theories in 6D, and for chiral or non-chiral supergravity
theories in four dimensions with extended supersymmetry, there should
be similar constraints on the set of F-theory constructions, which may
be a useful guide in discovering new constraints on which low-energy
field theories can consistently be coupled to quantum gravity in four
or six dimensions.  We hope that the work presented here will play a
useful role in leading to developments in this direction.
\vspace*{0.1in}

{\bf Acknowledgements}: We would like to thank Allan Adams, Maissam Barkeshli, Michael
Douglas, Antonella Grassi, Sheldon Katz, R. Loganayagam, 
and John McGreevy for helpful discussions.  DRM and WT would
like to thank the Aspen Center for Physics for hospitality in the
early stages of this work.  This research was supported by the DOE
under contract \#DE-FC02-94ER40818 and by the National Science Foundation
under grant DMS-0606578.  Any opinions, findings, and conclusions or recommendations expressed in this 
material are those of the authors
and do not necessarily reflect the views of the granting agencies.

\end{document}